\documentclass[12pt,prx,aps, reprint,showpacs]{revtex4-2}

\usepackage{graphicx}
\usepackage{amsfonts}
\usepackage{amsmath}
\usepackage{mathtools}
\usepackage{float}
\usepackage{epsfig}
\usepackage{color}
\usepackage{etoolbox}
\usepackage{gensymb}
\usepackage{multirow}
\usepackage{subcaption}
\usepackage[mathscr]{euscript}
\usepackage{pifont}

\newcommand{\e}[1]{\ensuremath{\times 10^{#1}}}
\newcommand{\ee}[1]{\ensuremath{10^{#1}}}
\newcommand{\dtwomin}[0]{\ensuremath{D^2_{\mbox{min}}}}

\usepackage{empheq}
\usepackage{physics}
\newcommand{\ignore}[1]{}

\begin{document}

\title{Interplay of rearrangements, strain, and local structure during avalanche propagation}

\author{Ge Zhang}


\affiliation{\emph{Department of Physics and Astronomy}, \emph{University of Pennsylvania},
Philadelphia PA 19104 }

\author{Sean Ridout}

\affiliation{\emph{Department of Physics and Astronomy}, \emph{University of Pennsylvania},
Philadelphia PA 19104 }

\author{Andrea J. Liu}

\email{ajliu@physics.upenn.edu}

\affiliation{\emph{Department of Physics and Astronomy}, \emph{University of Pennsylvania},
Philadelphia PA 19104 }

\begin{abstract}

Jammed soft disks exhibit avalanches of particle rearrangements under quasistatic shear. We follow the avalanches using steepest descent to decompose them into individual localized rearrangements. We characterize the local structural environment of each particle by a machine-learned quantity, softness, designed to be highly correlated with rearrangements, and analyze the interplay between softness, rearrangements and strain. Our findings form the foundation of an augmented elastoplastic model that includes local structure.

\end{abstract}

\maketitle
\section{Introduction}
All disordered solids respond elastically at low strain but flow plastically at sufficiently high strain. As strain increases beyond the elastic regime, disordered solids partially relax via intermittent localized rearrangements until they reach the yield strain, where they begin to flow. Up to the yield strain, disordered solids display surprisingly universal behavior~\cite{cubuk2018Science}. Beyond the yield strain, however, disordered solids exhibit several different classes of plastic behavior.  Foams can flow indefinitely via localized rearrangements without ever fracturing \cite{durian1995foam}.  Many systems exhibit crackling noise or avalanche behavior \cite{sethna2001crackling,saljeDahmen2014crackling,sethna2017deformation,dahmen2019FrontiersinPhysicsreview}, while still others exhibit shear banding and brittle fracture \cite{conner2003shear}. Here we focus on avalanche behavior. 

An avalanche consists of a series of rearrangements. A class of models known as elastoplastic models describes the courses of avalanches in terms of the interplay of rearrangements and elastic stress~\cite{nicolas2017deformation}: an increase of elastic stress can cause a local region to yield and rearrange, while conversely, a local rearrangement can increase stress elsewhere. It has become increasingly clear, however, that rearrangements and elasticity do not tell the whole story. Systems with identical microscopic interactions can show ductile or brittle behavior depending on preparation history~\cite{shavitriggleman2014PhysChemChemPhys,ozawa2018pnas}. This bolsters approaches that postulate local structures prone to rearrange ~\cite{spaepen1977microscopic,falk2011review}, but also points to the need for microscopic understanding of the connection between local structure and the physics included in elasto-plasticity models. While it has been shown that certain local structural environments are much more likely to rearrange than others~\cite{harrowell2008irreversible,manning2011vibrational, schoenholz2016structural}, the effects of rearrangements on local structure have not been established, even though it is clear that they must exist. It is also clear that elastic stresses can distort the structural environment surrounding a particle.  These considerations point to the need for detailed understanding of the interplay of local structure, rearrangements and elasticity.  

\begin{figure*}
\includegraphics[width=\textwidth]{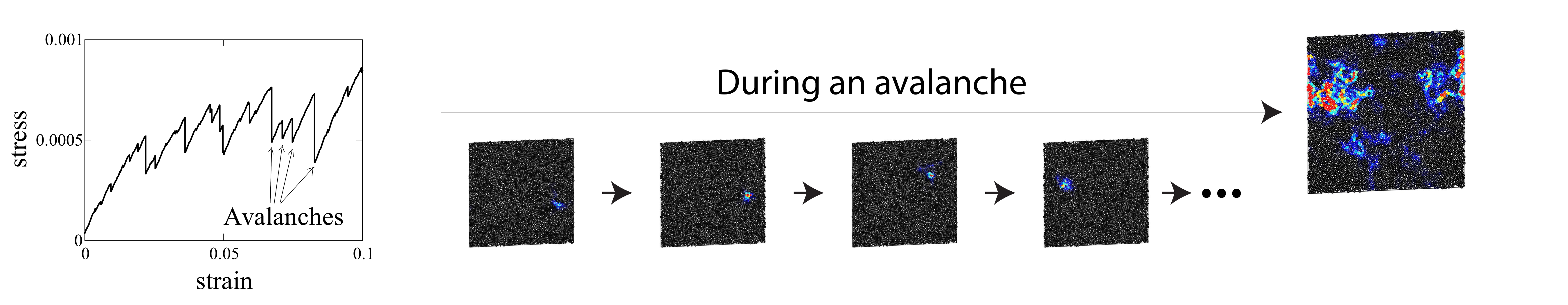}
\caption{As strain increases, avalanches occur during stress drops. During an avalanche, some constituent particles rearrange, triggering other localized rearrangements far away in the depicted system of $N=4000$ particles. Here, the non-affine displacement $\dtwomin$ of particles is represented on a black-to-blue-to-red scale with red corresponding to high values of $\dtwomin$. The rightmost plot depicts the cumulative $\dtwomin$ measured over the entire stress drop.}
\label{fig:qsshear}
\end{figure*}

In this paper, we go back to basics to untangle the interplay of local structure, rearrangements and strain in athermal, quasistatically sheared jammed packings of soft disks. We capture local structure via a machine-learned quantity, softness~\cite{schoenholz2016structural, cubuk2018Science, schoenholz2017relationship,sharp2018machine}. This quantity has been shown to provide useful insight into the dynamics of supercooled liquids and glasses~\cite{schoenholz2016structural,schoenholz2017relationship,sussman2018films}. Following this earlier approach~\cite{schoenholz2016structural}, we describe softness as the weighted sum of a set of structural quantities based on the local pair correlation function, where the weights are chosen to maximize the correlation with rearrangements that occur during avalanches. To tease out the interplay between softness, rearrangements and stress, we look at the effects of each of these factors on the others to develop a ``structuro-elasto-plasticity" framework for avalanches in disordered solids.

\section{Simulation details and softness}

We generate two-dimensional packings of $N$ soft disks in a simulation box with periodic boundary conditions. The disks interact with each other through the pairwise additive Hertzian potential:
\begin{equation}
u_2(r)=
\begin{cases}
\displaystyle \left( 1-\frac{r}{\sigma_i+\sigma_j} \right)^{2.5} ,& \text{if $r<\sigma_i+\sigma_j$,} \\
0,& \text{otherwise,}
\end{cases}
\label{function_form}
\end{equation}
where $\sigma_i$ is the radius of the $i$th disk. 
To avoid crystallization, we use a $1:1$ mixture of particles with $\sigma=0.5$ and $\sigma=0.7$.

Starting from random initial conditions, we minimize the potential energy to find the initial zero-temperature jammed state. We then repeatedly apply a small shear-strain step of $\delta \epsilon$, minimizing the energy after each step, until the total strain reaches $\epsilon_{end}$. 
The stress-strain relation for a single configuration, shown in Fig.~\ref{fig:qsshear}, confirms the existence of avalanches.
We generated 5 trajectories with $N=\ee{5}$, $\delta \epsilon=\ee{-5}$, and $\epsilon_{end}=0.1$; and 20 trajectories with $N=4000$, $\delta \epsilon=\ee{-4}$, and $\epsilon_{end}=2$. This smaller system with $N=4000$ is shown in Fig.~\ref{fig:qsshear} for visual clarity. It is also used to train the machine-learning algorithm because we need to access larger shear strains, as detailed in the supplementary information (SI)~\cite{supplement}. All of the remaining analysis was carried out on the larger system.

It is well known~\cite{maloney2006amorphous} that during athermal quasistatic shear, energy drops mark rearrangements that can be either localized or extended due to avalanches. In each step of strain followed by energy minimization, we calculate the final energy to monitor for energy drops. We focus only on energy drops, using steepest descent with adjustable step sizes~\cite{supplement} to capture the over-damped relaxation process from the beginning of the energy drop to the end. We save intermediate configurations that are equidistant in configuration space, so that the sum over particles of particle displacement squared is fixed between successive frames.

To identify rearranging particles, or ``rearrangers," we calculate $\dtwomin$ \cite{falk1998dynamics}:
\begin{equation}
\dtwomin(k)=\frac{1}{M_k}\sum_{i}^{M_k}\left[ \mathbf r^{'}_{ik} - \mathbf J_k \mathbf r_{ik} \right ]^2
\label{eq:d2min}
\end{equation}
where the sum is over all neighbors of particle $k$ within a distance of $R_{D}=2$. Here $M_k$ is the number of such neighbors, $\mathbf r_{ik}$ and $\mathbf r^{'}_{ik}$ are the vector separations between particles $i$ and $k$ at two consecutive frames, respectively, and $\mathbf J_k$ is the ``best-fit'' local deformation gradient tensor about particle $k$ that minimizes $\dtwomin$. A particle with $\dtwomin$ above a certain threshold~\cite{supplement} is a rearranger.
The rest of the paper presents results for rearrangers that are small particles in our binary mixture, but we have verified that results for large-particle rearrangers are qualitatively the same.

\begin{figure*}
\begingroup
\setlength{\tabcolsep}{0.1pt} 
\renewcommand{\arraystretch}{0.1} 
\begin{tabular}{ccc}
\includegraphics[width=0.33\textwidth]{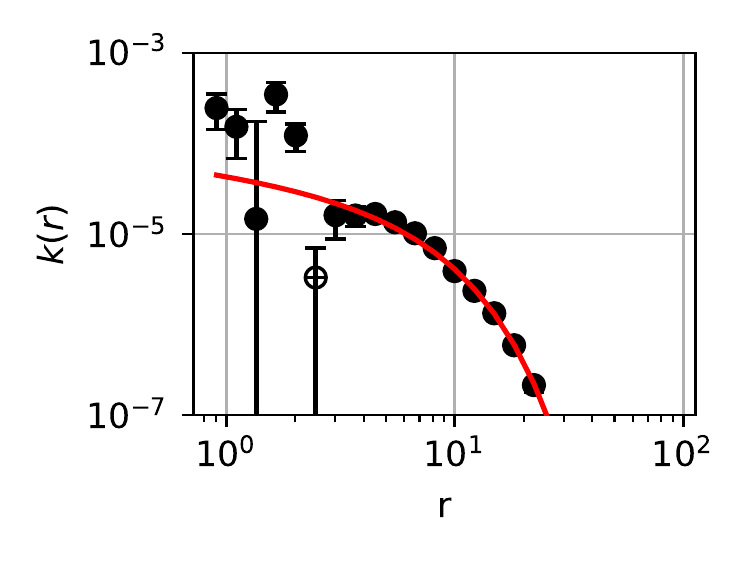}&
\includegraphics[width=0.33\textwidth]{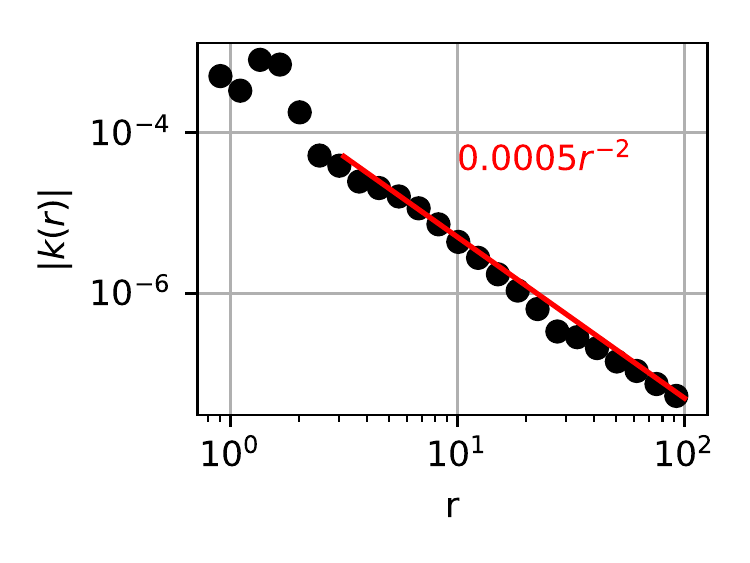}&
\includegraphics[width=0.33\textwidth]{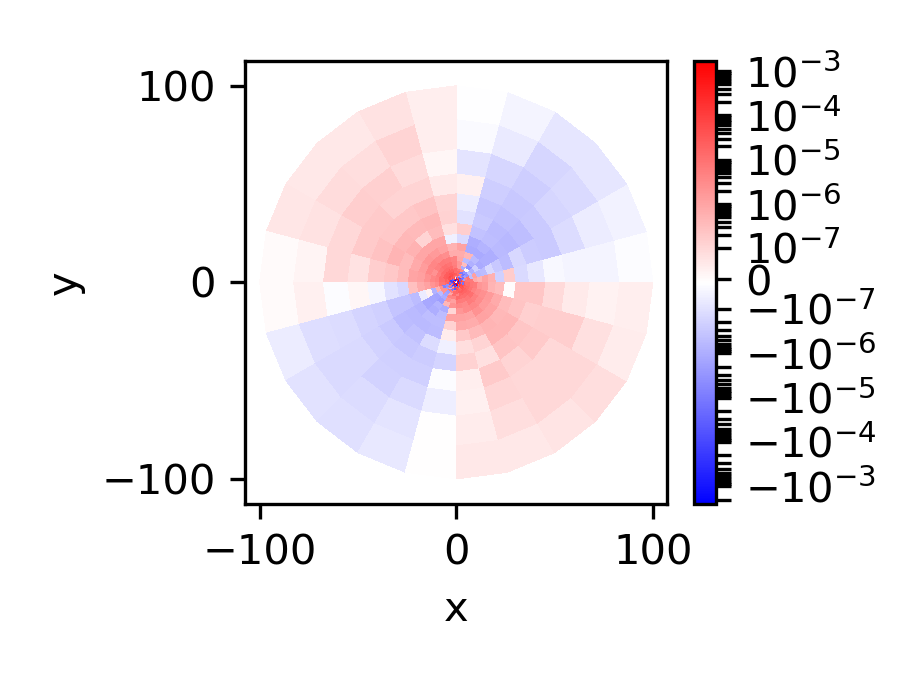}\\
\includegraphics[width=0.33\textwidth]{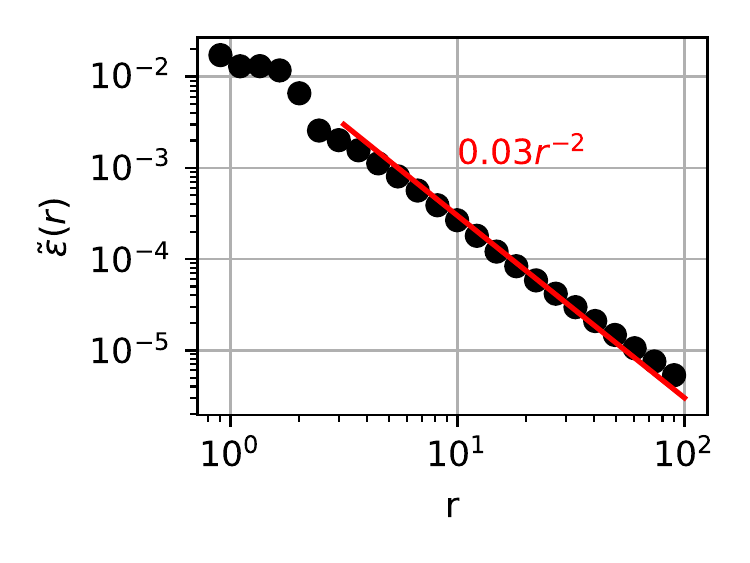}&
&
\includegraphics[width=0.33\textwidth]{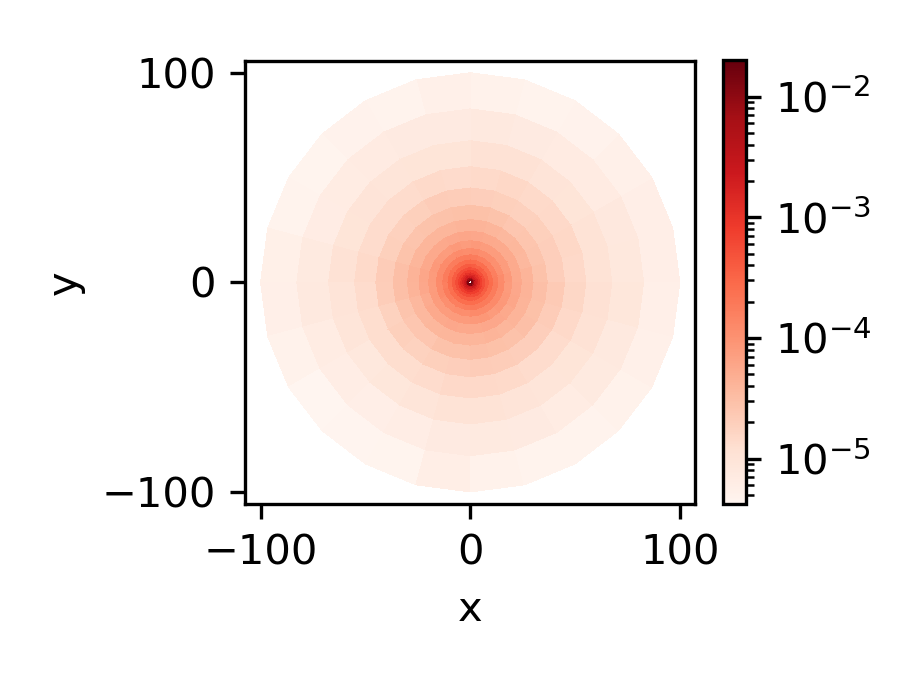}\\
\includegraphics[width=0.33\textwidth]{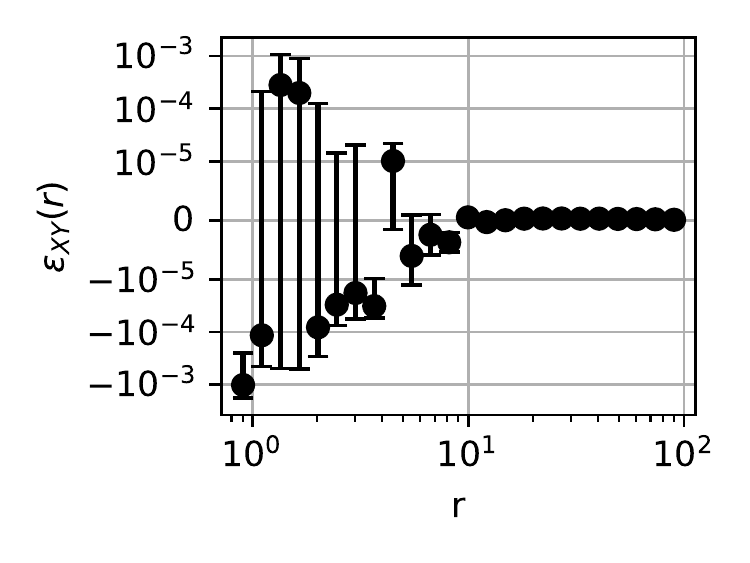}&
\includegraphics[width=0.33\textwidth]{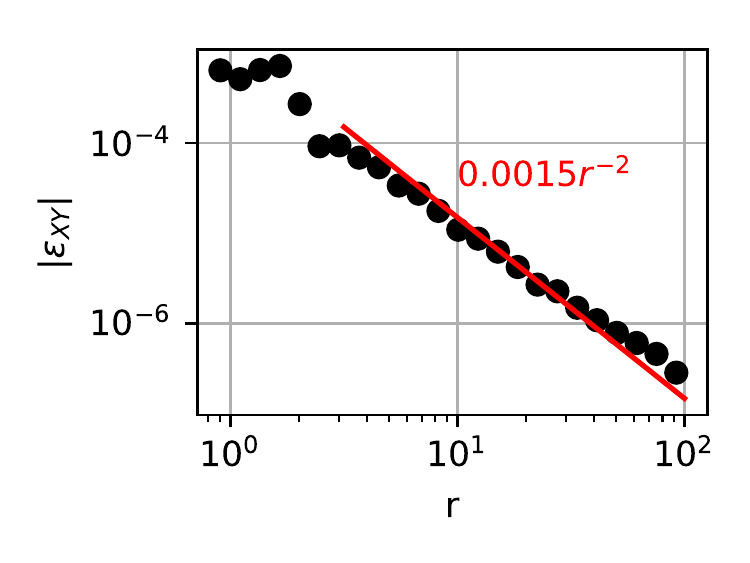}&
\includegraphics[width=0.33\textwidth]{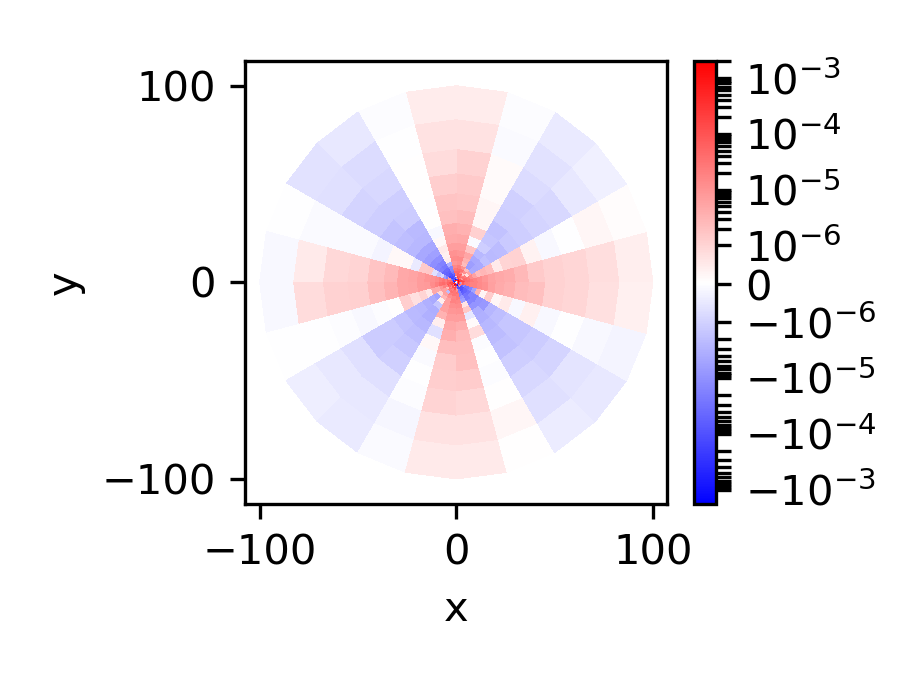}
\end{tabular}
\endgroup
\caption{Mean volumetric strain $k$ (top row), mean deviatoric strain ${\tilde \epsilon}$ (middle), and mean shear strain in the $xy$ direction (the direction of the global shear) $\epsilon_{xy}$ (bottom) per frame caused by a rearranging particle at the origin. 
Angular-averaged (left column), angular-averaged absolute value (middle), and angular (right) versions are shown.
Note that the middle row/column plot is not shown because ${\tilde \epsilon}$ is always non-negative.
In the top left plot, solid circles represent positive values of $k(r)$, while open circles represent negative values.
Red lines are fits to continuum-elasticity predictions detailed in the text and Appendix.
}
\label{fig:strainField}
\end{figure*}

We draw from the sets of rearrangers and non-rearrangers to train a linear support vector machine to obtain softness, a structural quantity that indicates the propensity of a particle to rearrange \cite{schoenholz2016structural}. As detailed in the SI~\cite{supplement}, the softness of particle $i$ is essentially a weighted integral over the local pair correlation function $g_i(r)$.
The weight function is inferred by the linear support vector machine to maximize the accuracy of predicting rearrangers. As in Ref.~\cite{schoenholz2016structural}, the weighting is highly negative at the first peak of $g(r)$, implying that particles with fewer neighbors have higher softness, consistent with intuition based on the cage picture.

\section{The avalanche process}

In Fig.~\ref{fig:qsshear} and the supplemental video \cite{supplement}, we confirm that during avalanches, rearrangements are indeed localized and sequential, as assumed in elastoplastic models~\cite{nicolas2017deformation}. 
Moreover, consecutive rearrangements can be very far apart.

\begin{figure}
\begin{subfigure}{0.4\textwidth}
\includegraphics[width=\textwidth]{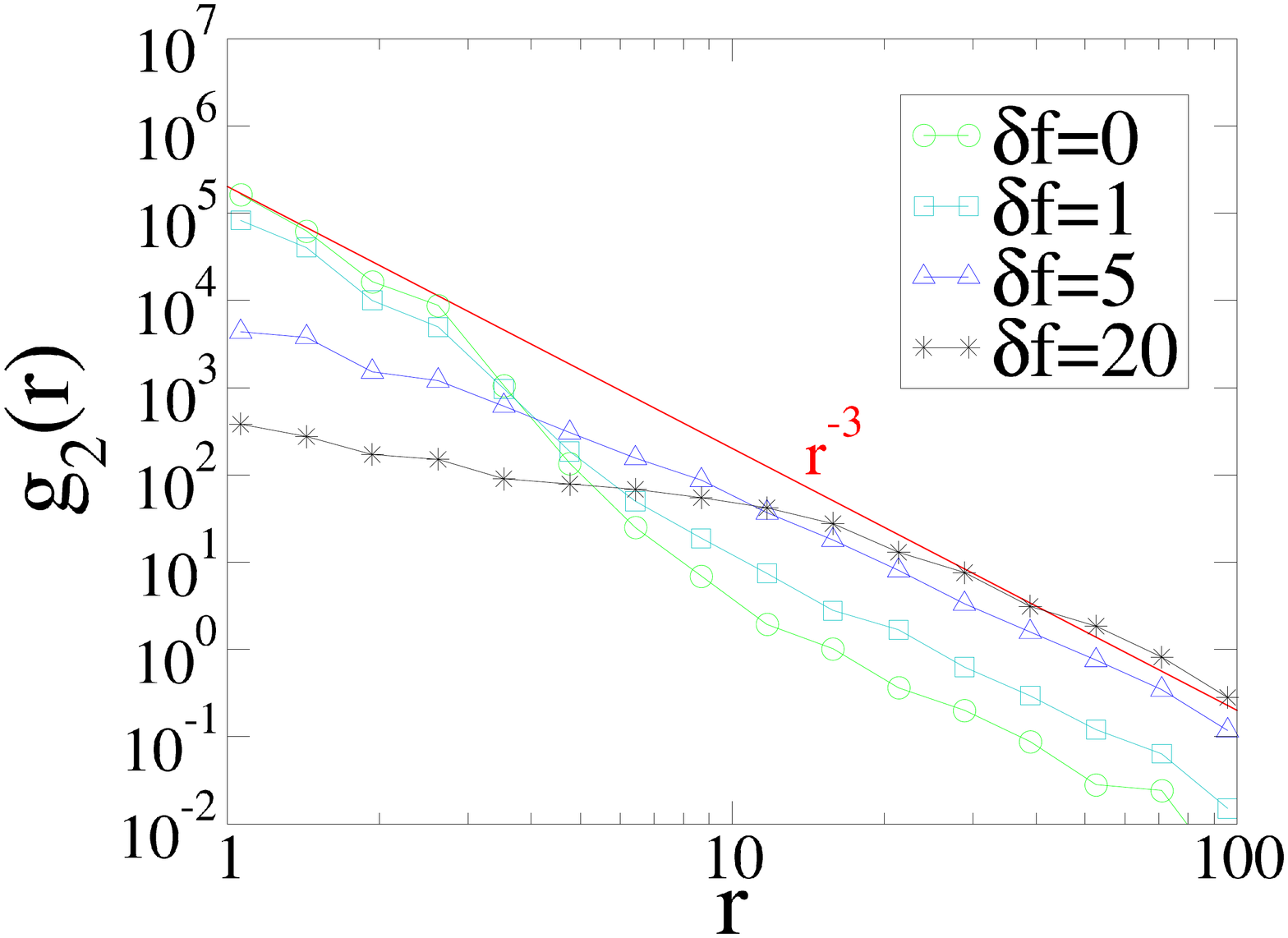}
\caption{}
\end{subfigure}
\begin{subfigure}{0.48\textwidth}
\includegraphics[width=\textwidth]{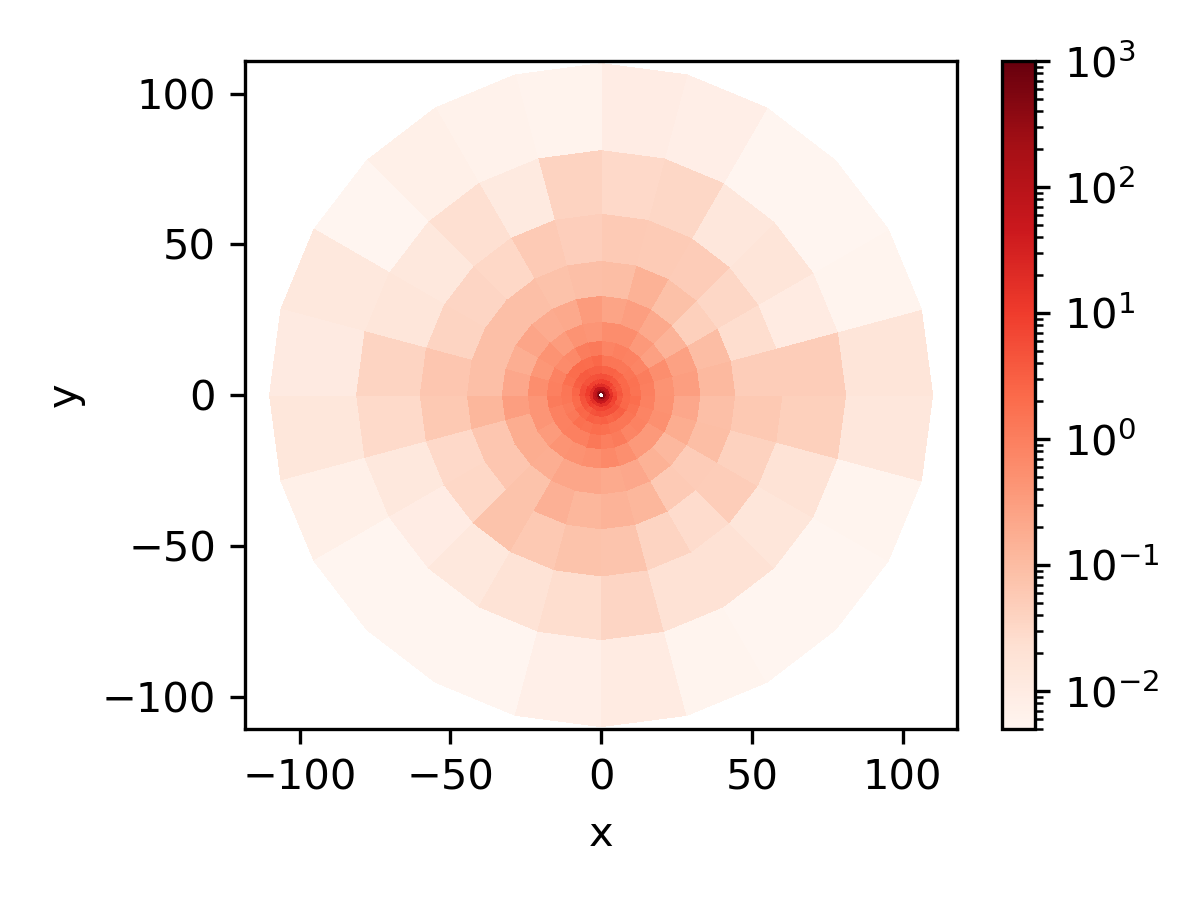}
\caption{}
\end{subfigure}
\caption{(a) The time-dependent pair correlation function of rearrangers, $g_2(r, \delta f)$ for different numbers of frames $\delta f$ following the rearrangement at the origin at frame $f=0$.  (b) The time-averaged directional plot $g_2(\mathbf r)=\frac{1}{F}\sum_{\delta f=0}^{F} g_2(\mathbf r, \delta f)$, where $F=20$.}
\label{fig:rearrangerCorrelation}
\end{figure}

\subsection{Interplay of rearrangements with strain}
We begin our study of the interplay between rearrangements, softness and strain by examining the effect of a rearrangement at the origin on strain at $\bf r$, averaged over many rearrangements. The local-fit deformation tensor about each particle, $\mathbf J$, is already known from calculating $\dtwomin$. From $\mathbf J$, which is a second-rank tensor with four degrees of freedom in 2D, we extract three different strain components: the volumetric (isotropic) strain $k$, total deviatoric strain ${\tilde \epsilon}$, and shear strain in the $xy$ direction  (the direction of the global shear), $\epsilon_{xy}$. We extract these by comparing two consecutive frames when a rearrangement is occurring (see SI).

The near-field behaviors of the local strains depend on microscopic details of how rearrangements locally strain their surroundings, but in the far field we expect the local strains to be well-described by elasticity theory. In the far field, one typically approximates the rearrangement as a point plastic shear strain, equivalent to a pair of point force dipoles. The responses to this source of the three local strains studied at position $\mathbf r$ and time $t$ following a rearrangement at the origin at $t=0$ are given in Eqs.~(\ref{eq:analyticalK})-(\ref{eq:analyticalExx}). Because the actual shear strain source due to the rearrangement is very long-lived, however, the response to the point plastic shear strain is well-approximated by the infinite-time limit, shown in Eq.~(\ref{eq:epsilon}). 

\begin{figure}
\begin{subfigure}{0.48\textwidth}
\includegraphics[width=\textwidth]{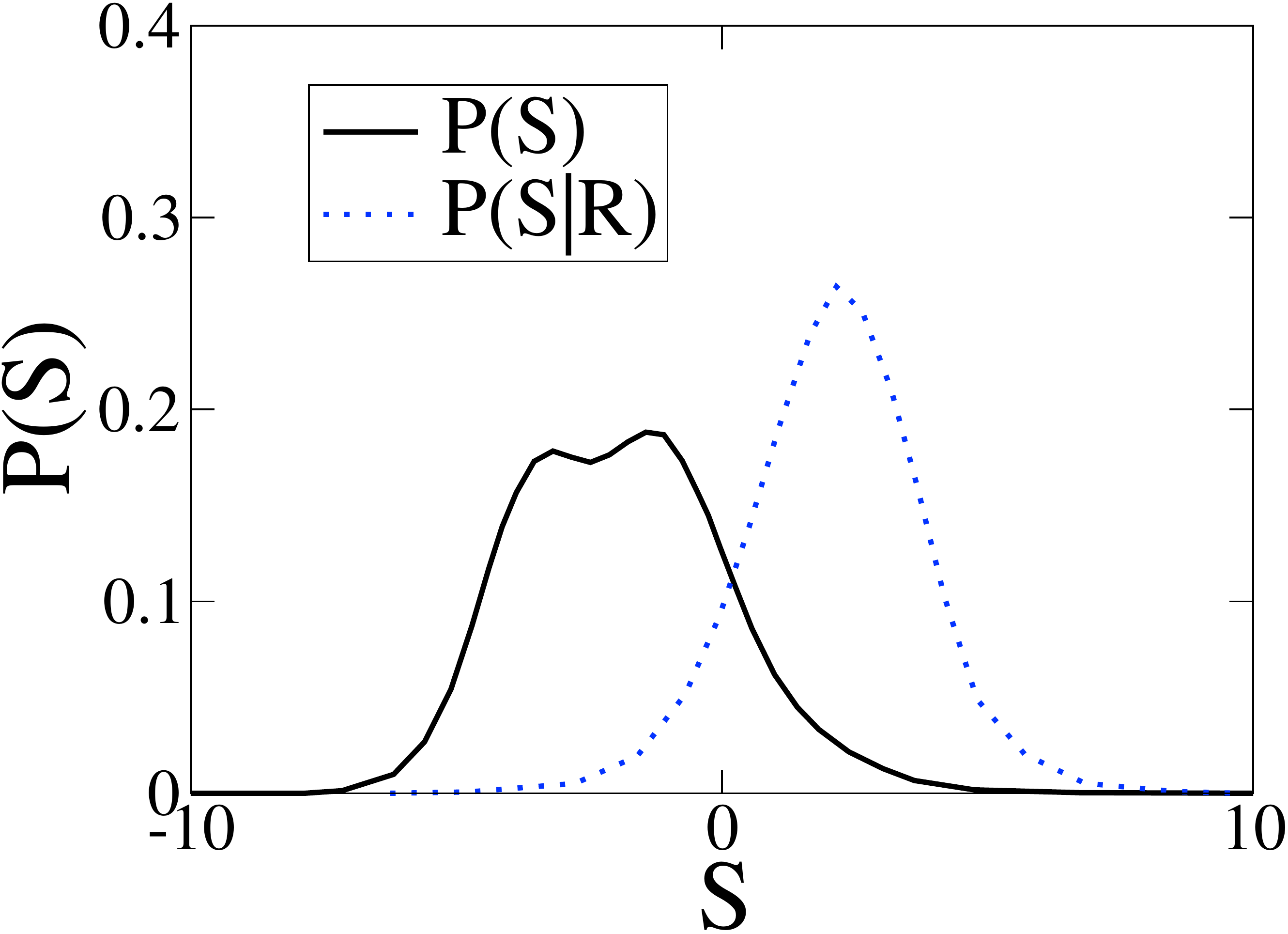}
\caption{}
\end{subfigure}
\begin{subfigure}{0.48\textwidth}
\includegraphics[width=\textwidth]{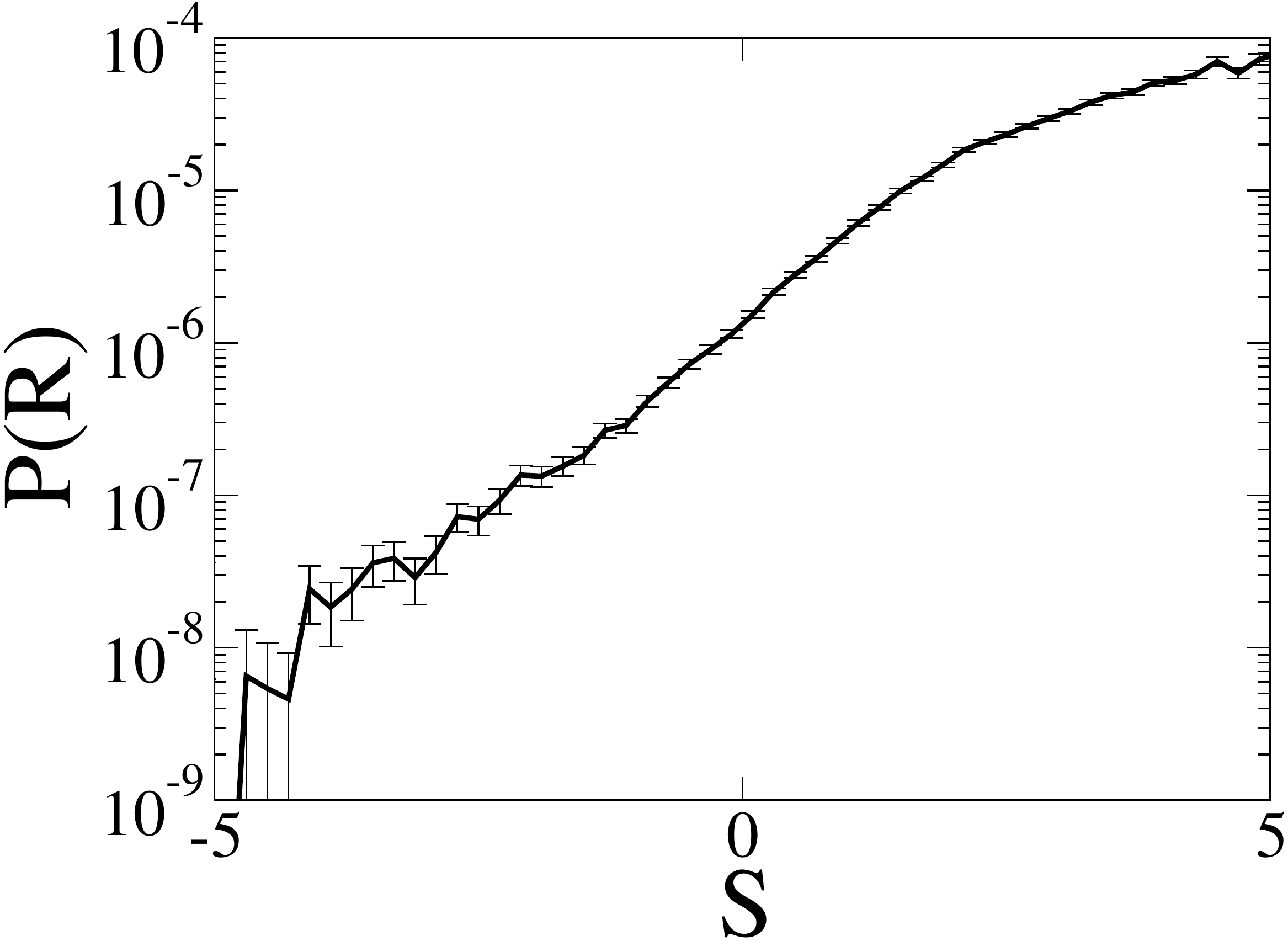}
\caption{}
\end{subfigure}
\caption{Performance of machine-learned softness. (a) The distribution of softness for all particles (black solid) and for rearrangers only (blue dotted). There is a pronounced difference between the two distributions. (b) The probability that a particle is rearranging, $P_R$, as a function of its softness. As the softness increases, $P_R$ increases by four orders of magnitude, verifying the high correlation between softness and rearrangements.
}
\label{fig:softnessDistribution}
\end{figure}

Analytical derivations \cite{picard2004elastic}, numerical measurements \cite{maloney2006amorphous}, and experiments \cite{jensen2014local} have all found that $\epsilon_{xy}$ has an $r^{-2}$ radial dependence and a quadrupolar angular dependence in response to a point force dipole. Indeed our results show that $|\epsilon_{xy}|$ decays as a power law close to $r^{-2}$ with a quadrupolar angular dependence, as indicated by the bottom row of Fig.~\ref{fig:strainField}. The red solid lines are fits to the dependence expected from continuum elasticity (see Appendix A). Because of the quadrupolar angular dependence, the angular average of $\epsilon_{xy}(r)$ is zero within statistical noise (almost every error bar crosses the $x$ axis). 
 The deviatoric strain, ${\tilde \epsilon}$(middle row in Fig.~\ref{fig:strainField}), likewise decays as $r^{-2}$ (red solid line in left plot) but with an isotropic angular dependence (right plot), as expected from continuum elasticity (see Appendix A). 

The volumetric strain $k(r)$ is typically neglected in systems of fixed total volume but as we will show, it plays an important role because softness is strongly dependent on local density. The volumetric strain in response to a shear strain is given in Eq.~(\ref{eq:analyticalK}). 
The far field response to a rearrangement, however, must also take into account the effect of a point compression source since the rearrangement can also give rise to local plastic compression. This has a transient effect since the total volume of the system is conserved, but is significant because it gives rise to a contribution to $k(r)$ [Eq.~(\ref{eq:kr})] that does not angle-average to zero. As a result,
the volumetric strain $k(r)$ is predicted by elasticity theory to be the sum of two terms: a $\sin(2\theta)r^{-2}$ term arising from the point shear strain and another term arising from the point compression. 
The top left plot of Fig.~\ref{fig:strainField} shows that the angular-averaged volumetric strain $k(r)$ is positive at most $r$ and does not exhibit a power-law decay; this corresponds to the second term.  In the Appendix we discuss the expectation from elasticity theory that yields the fit (red solid curve) shown. The absolute value of $k(r)$ is dominated by the first term and shows an $r^{-2}$ decay, consistent with continuum elasticity theory for a point plastic shear strain (red solid line in top middle plot).
For $r \gtrsim 5$, $k$ has the expected dipolar angular dependence from the first term (top right plot). In summary, of the two contributions to the local volumetric strain, the term arising from the point shear strain dominates but angle-averages to zero so that the second term is revealed in the angular-averaged $k(r)$.

Although the results shown here are for two-dimensional systems, we have confirmed that the expected scalings for volumetric and deviatoric strain are observed in 3 dimensions~\cite{supplement}, providing strong evidence in favor of our interpretation of the roles of volumetric, deviatoric and $xy$-strain.  

We now turn to the effect of the induced strain on the next rearrangement. We first compute the frame-dependent pair correlation function of rearrangers $g_2(\mathbf r, \delta f)$, namely the probability of finding a rearrangement at $\mathbf r$ $\delta f$ frames later, given a rearrangement at the origin at frame $\delta f=0$. Results for several values of $\delta f$ are presented in Fig.~\ref{fig:rearrangerCorrelation}.

\begin{figure*}
\begin{subfigure}{0.38\textwidth}
\includegraphics[width=\textwidth]{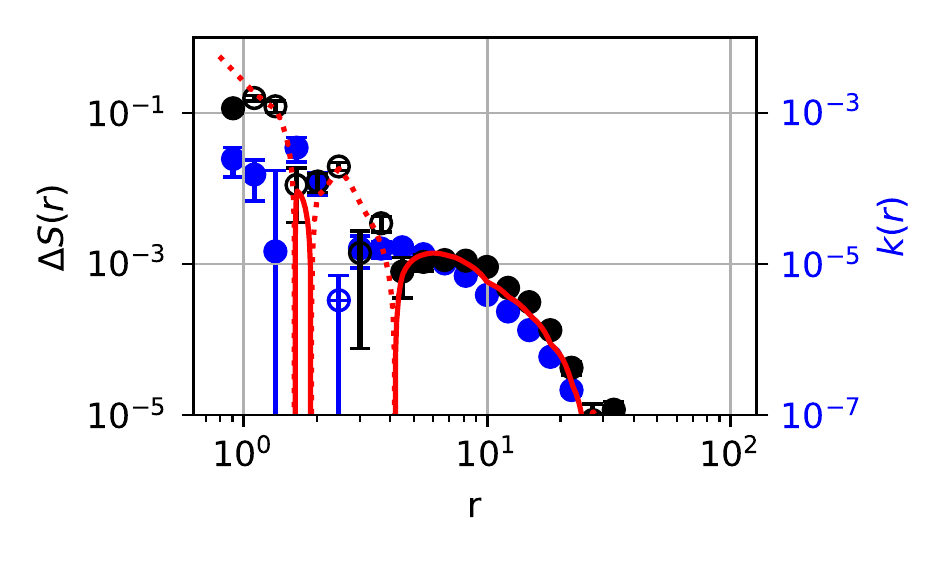}
\caption{}
\end{subfigure}
\begin{subfigure}{0.3\textwidth}
\includegraphics[width=\textwidth]{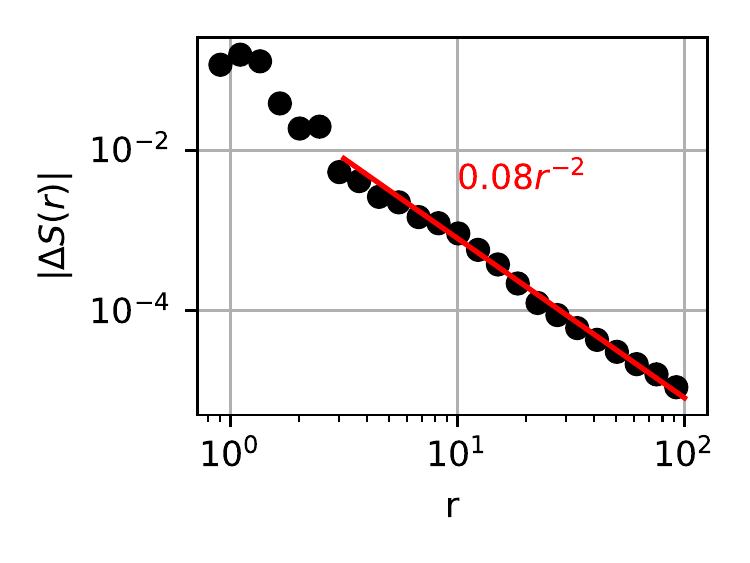}
\caption{}
\end{subfigure}
\begin{subfigure}{0.3\textwidth}
\includegraphics[width=\textwidth]{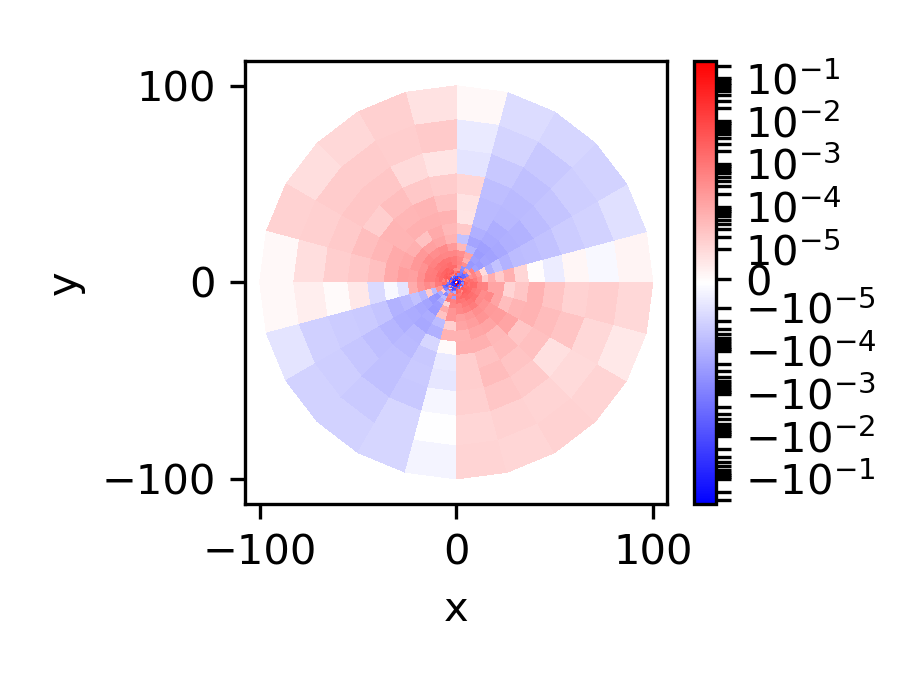}
\caption{}
\end{subfigure}
\caption{(a) Mean softness change, $\Delta S$, per frame caused by a rearranging particle at the origin. A prediction from Eq.(\ref{eq:softnessDecompose}) is plotted as red lines. Here we also plot the volumetric strain $k(r)$ for comparison. 
Similar to Fig.~\ref{fig:strainField}, solid dots and solid lines represent positive values, while hollow circles and dotted lines represent negative values.
(b) Same as (a), but for its absolute value.
(c) Mean softness change with directional dependence shown.}
\label{fig:rearranger}
\end{figure*}

We first focus on the radial dependence. The rearranger pair correlation function $g_2(r, \delta f)$ decays as $r^{-3}$, independent of $\delta f$. This is consistent with either ${\tilde \epsilon}$ or $\epsilon_{xy}$, which both decay as $r^{-2}$, due to the following argument. Two earlier studies of systems with spherically-symmetric potentials found that the cumulative distribution of the local yield strain has a low-yield-strain tail described by a power law with exponent 1.6~\cite{karmakar2010,barbot2018local}.
On general grounds this scaling should also apply to our system~\cite{karmakar2010}, so the probability that a rearrangement is triggered by $\tilde \epsilon$ or $\epsilon_{xy}\sim r^{-2}$ should scale as $(r^{-2})^{1.6}=r^{-3.2}$, roughly consistent with the scaling we observe in $g_2$.

\begin{figure}
\begin{subfigure}{0.48\textwidth}
\includegraphics[width=\textwidth]{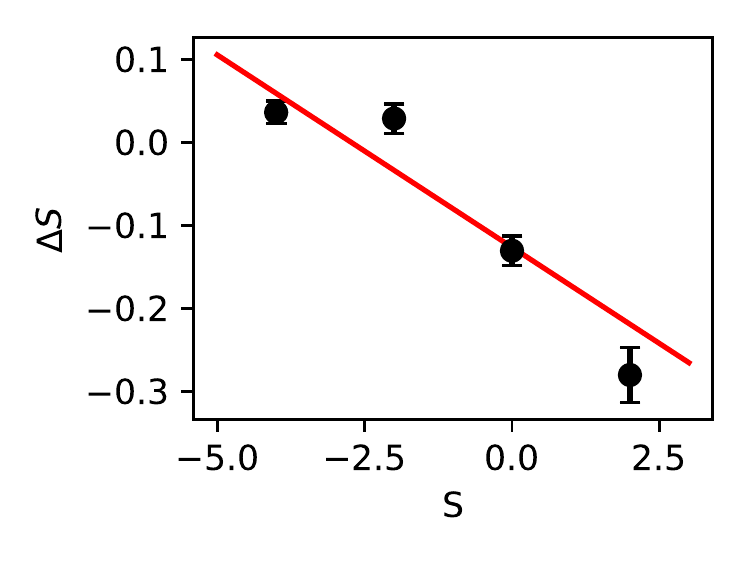}
\caption{}
\end{subfigure}
\begin{subfigure}{0.48\textwidth}
\includegraphics[width=\textwidth]{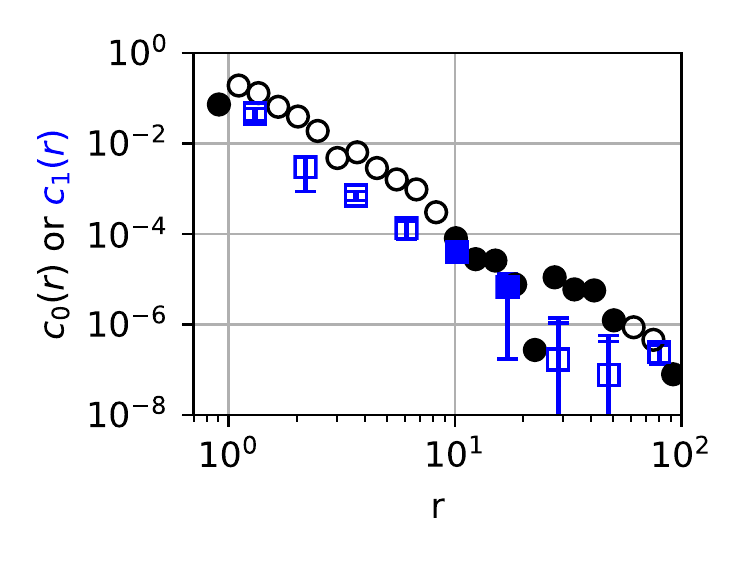}
\caption{}
\end{subfigure}
\caption{(a) Mean softness change, $\Delta S$, per frame for a particle with a given $S$ within a distance of $r=1.6$ of a rearranger. The red line is the linear fit. (b) The slope of such linear fits, $c_1$ (squares), as well as $c_0$ (circles) defined in Eq.~(\ref{eq:softnessDecompose}), at different distances $r$. Solid symbols represent positive values, while open ones  represent negative values.}
\label{fig:softnessFeedback}
\end{figure}
Note that the angular dependence of $g_2(\mathbf r)$ is nearly isotropic and clearly does not show a quadrupolar dependence. This is consistent with the angular dependence of $\tilde \epsilon$, not $\epsilon_{xy}$ (see Fig.~\ref{fig:strainField}). We therefore conclude that rearrangement-induced shear strain in any direction can trigger rearrangements equally well. This result contradicts the assumption of many elastoplastic models that $\epsilon_{xy}$ is solely responsible for triggering rearrangements.

\subsection{Interplay of softness with rearrangements and strain}

In training the machine-learning algorithm to obtain softness, we find that 90\% of rearrangers have $S>0$, while 84\% of non-rearrangers have $S<0$. 
Moreover,
Fig.~\ref{fig:softnessDistribution} shows that the softness distribution for rearrangers is very different from that of the whole population, and that the probability that a particle rearranges increases by four orders of magnitude as softness increases. These results verify that softness affects the propensity to rearrange very strongly. 

\begin{figure}
\includegraphics[width=0.48\textwidth]{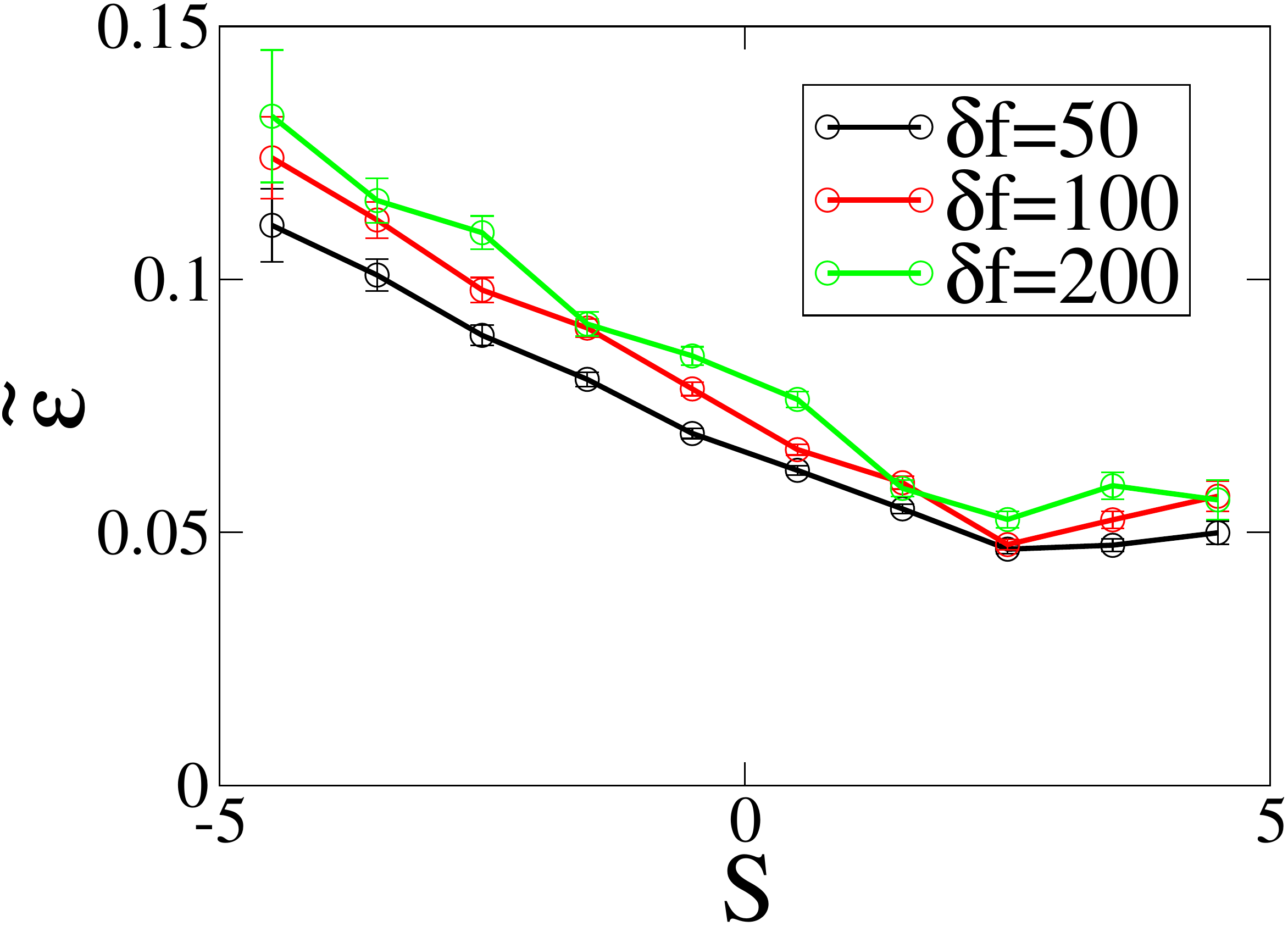}
\caption{The amount of shear strain exerted to the local environment of a particle before it starts to rearrange versus the softness of that particle, observed 50, 100, and 200 frames before the rearrangement.}
\label{fig:shearStrainToRearrange}
\end{figure}

In turn, rearrangements can affect softness. The average difference in softness of a rearranger immediately before and after the rearrangement is $\langle \Delta S \rangle_R = -0.75$; the softness of a rearranger drops significantly when it rearranges. Rearrangements can also affect the softness of other particles;   
we plot the mean softness change $\Delta S (\bf r)$ of a particle at $\bf r$ due to a rearrangement at the origin in Fig.~\ref{fig:rearranger}. Rearrangements make directly contacting neighbors ($r<1$) softer and non-contacting nearby particles ($1<r<5$) less soft. Rearrangements also make far-away particles ($r>5$) softer or harder depending on the orientation. The distance and angular dependences of the far-field $\Delta S$ are consistent with the volumetric strain $k$ (see Fig.~\ref{fig:strainField}), suggesting that it is caused by $k$. This is not surprising since softness is sensitive to density.

To understand the near-field effect of rearrangements on softness, we first note that in a thermal Lennard-Jones system, the mean softness of non-rearranging particles with a given initial softness $S_0$ evolves toward the global mean value for any $S_0$ \cite{schoenholz2016structural} due to rearrangements of other particles. In other words, rearrangements of other particles tend to push softness towards its mean value. Here we study if the same effect exists in our quasistatically sheared system. For particles within a short distance $r\le1.6$ to a rearranger, we plot the softness change vs. the original softness and perform a linear fit, presented in Fig.~\ref{fig:softnessFeedback} (a). We plot the slopes $c_1(r)$ of such fits at several different $r$ in Fig.~\ref{fig:softnessFeedback} (b). For $r<10$ and $r>30$, $c_1$ is negative, indicating that softness in our system also has the tendency to approach its mean at these distances. However, $c_1$ is positive for $10<r<30$, suggesting the opposite effect. The effect is small and negligible, and is probably because softness tend to increase in this range of $r$ [see Fig.~\ref{fig:rearranger} (a)], and the softer a particle is, the floppier its local environment is, and the more tendency it has to deform, even if such deformation generally raises $S$. More important is the magnitude of $c_1(r)$: we see that the magnitude of $c_1(r)$ decays rapidly with $r$ and is well described as a power law: $\abs{c_1(r)}=0.06r^{-3.2}$. Finally, $c_1(r)$ appears to be independent of the angle $\theta$.

Overall, our results suggest that the mean softness change of a particle with softness $S$ at $\mathbf r$ when a particle at the origin rearranges is:
\begin{equation}
    \Delta S(\mathbf r, S)= c_0(r)+c_1(r)(S-\langle S \rangle)+b k(\mathbf r)
\label{eq:softnessDecompose}
\end{equation}
where $c_1(r)$ is given in Fig.~\ref{fig:softnessFeedback} (b), and $b\approx 207$. To find $c_0$, we subtract $b k(\mathbf r)$ from $\Delta S(\mathbf r)$. Similar to $c_1$, we do not find any angular dependence in $c_0$. We plot its $r$-dependence in Fig.~\ref{fig:softnessFeedback}(b). Clearly, $c_0$ and $c_1$ exhibit similar power law decays; we find $\abs{c_0(r)}=0.3 r^{-3.1}$. With the fit, Eq.~(\ref{eq:softnessDecompose}) yields the red curve in Fig.~\ref{fig:rearranger}(a). 
Note that the red curve provides an excellent description of the black points ($\Delta S(r)$), capturing the sign as well as the magnitude except for two points at small $r$.

\subsection{How strain and softness induce new rearrangements}
We have shown that rearrangements give rise to deviatoric strain that in turn triggers new rearrangements. We have also shown that rearrangers tend to have high softness. Here we study how $S$ and ${\tilde \epsilon}$ work in tandem to induce rearrangements. When a particle starts rearranging at frame $f$, we rewind $\delta f$ frames to calculate the shear strain exerted on this particle between $f-\delta f$ and $f$, and the softness $S$ at frame $f-\delta f$. As Fig.~\ref{fig:shearStrainToRearrange} shows, the amount of shear strain needed to trigger a rearrangement depends strongly on $S$, but only very weakly on $\delta f$. Thus, softer particles require less shear strain to start rearranging. This is consistent with earlier results in thermal systems that found that softer particles have lower activation energies to rearrange \cite{schoenholz2016structural, sharp2018machine}.
Indeed, we have conducted thermal molecular dynamics simulations to find energy barriers comparable to those predicted by Fig.~\ref{fig:shearStrainToRearrange}~\cite{supplement}.

\section{Discussion}

\begin{figure}
\includegraphics[width=0.48\textwidth]{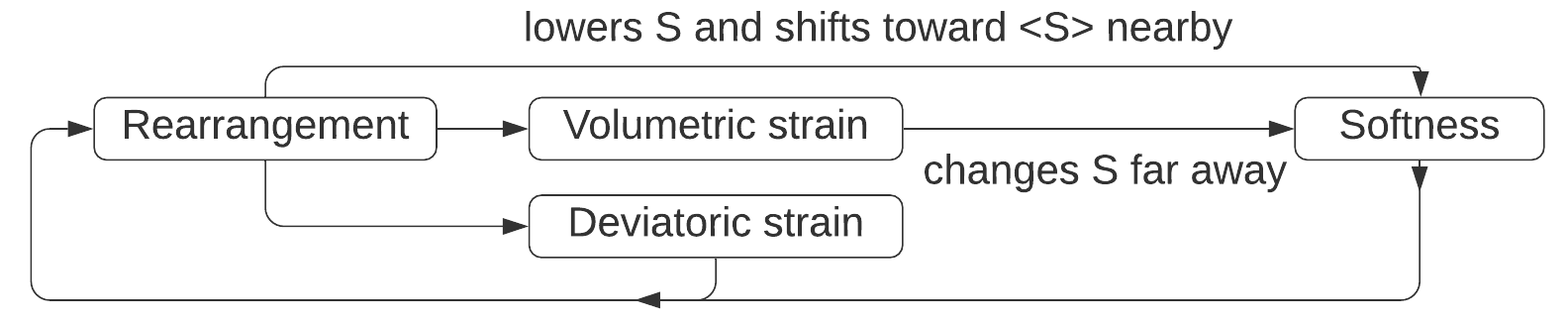}
\caption{Summary of the interplay between rearrangements, strains, and softness we identified. A rearrangement decreases the softness of nearby particles, alters the softness of far-away particles through volumetric strain, and exerts a deviatoric shear strain on all particles. The local deviatoric strain and softness together determine future rearrangements.}
\label{fig:summary}
\end{figure}

In this paper, we study avalanches that occur during energy drops when a jammed binary Hertzian disk packing is sheared quasistatically, using steepest descent to follow the minimization process. We have confirmed the consistency of our interpretation of the roles of volumetric and deviatoric strain in three-dimensional systems~\cite{supplement}. We expect that the qualitative results of Fig.~\ref{fig:summary} apply quite generally to both two and three-dimensional ductile disordered solids, which generically exhibit avalanche behavior.
We find that (1) a rearrangement alters the softness of a nearby particle according to the difference between its softness and the mean softness. This behavior was first observed for 3D Lennard-Jones systems above the glass transition~\cite{schoenholz2016structural}, indicating that it is quite general. (2) A rearrangement alters the softness of distant particles through volumetric strain. The existence of a transient volumetric strain, which has not been considered significant, is a feature of elasticity. The fact that local dilation/compaction increase/decrease the softness is consistent with the previously observed dependence of softness on local density in 3D Lennard-Jones mixtures and with our physical understanding of softness~\cite{schoenholz2016structural}, and is therefore also quite general. (3) A rearrangement exerts a deviatoric strain on the rest of the system. This should be generally true for isotropic systems in any dimension. (4) The average yield strain decreases with increasing softness. This is consistent with previous results for 3D Lennard-Jones simulations~\cite{schoenholz2016structural}, 2D colloidal glass experiments~\cite{ma2018prl} and 3D aluminum polycrystal simulations~\cite{sharp2018machine}, showing that the energy barrier for rearrangements decreases with increasing softness. 

Due to the generic nature of our findings, the summary of our results in Fig.~\ref{fig:summary} should hold generally for avalanches in ductile disordered solids.
Fig.~\ref{fig:summary} can be viewed as a structuro-elasto-plastic model that builds upon earlier elasto-plastic models. There are two main differences compared to the earlier models. First, we find that shear strain in any direction due to a rearrangement can trigger the next rearrangement equally well.  Elasto-plastic models typically focus on the component of the local shear strain with the same orientation as the global shear strain~\cite{budrikis2013avalanche,nicolas2017deformation}. Second, and more significantly, we have elucidated how the local structural environment of a particle affects and is affected by rearrangements and strain. 

Our results point to a few factors that may contribute to the ductile behavior observed. Future rearrangements are triggered by the total deviatoric strain rather than the $xy$-shear strain. As a result, rearrangements trigger successive rearrangements that are isotropically distributed, not concentrated in the direction of maximum $xy$-strain. In addition, a rearranger lowers the softness of nearby particles, discouraging them from rearranging, while on average raising the softness of distant particles, facilitating their rearrangement. Third, rearrangements tend to push the softness of nearby particles towards the mean, which is quite high for the ductile system. Our approach can be applied directly to systems that exhibit shear-banding and brittle failure to see whether the interplay is different in such systems. Earlier papers have shown that softness is readily identified in experimental systems for which the positions of particles can be tracked with time~\cite{cubuk2015PRL,cubuk2018Science,harrington2019pre}. Our analysis for disentangling the interplay of softness, rearrangements and strain can therefore be applied directly to experiments as well as simulations. It is likely that the key to understanding ductile vs. brittle behavior is encapsulated in this interplay.

\begin{acknowledgments}
We thank Hongyi Xiao, Robert J. S. Ivancic, Douglas J. Durian and Robert A. Riggleman for helpful discussions and their careful reading of the manuscript.
This work was supported by the UPenn MRSEC NSF-DMR-1720530 (GZ), the Simons Foundation for the collaboration
``Cracking the Glass Problem'' award $\#$454945 to AJL (GZ,SR,AJL), and Investigator award $\#$327939 (AJL), the U.S. Department of Energy, Office of Basic Energy Sciences, Division of Materials Sciences and Engineering under Award DE- FG02-05ER46199 (GZ), and NSERC via a PGS-D fellowship (SR).
\end{acknowledgments}

\appendix

\section{Continuum-elastic predictions for strain field induced by a rearrangement}
\label{sec:prediction}

The far field of rearrangement events has long been modelled as that of an Eshelby inclusion, which is the elastic response to a point strain source \cite{picard2004elastic}. 

Elastoplastic models typically only consider $\sigma_{xy}$, use an elastic kernel which assumes the medium to be incompressible and take the limit of infinite time (mechanical equilibrium). Since we are interested in understanding the course of avalanches during steepest descent, we need the kernel at finite times with overdamped dynamics. We sketch below the derivation of all components of the continuum strain field.

We begin by considering an infinite elastic medium subject to a point force turning on at $t=0$ at the origin.

We wish to find $G_{ik}{\left(\mathbf{r}, t\right)}$ such that

\begin{equation}
C_{jpim} \pdv[2]{G_{ik}}{x_p}{x_m} - \eta \pdv{G_{ik}}{t} + \delta_{jk} \delta{(\mathbf{r})} \Theta{\left(t\right)} = 0.
\end{equation}

Taking a Fourier transform in space and a Laplace transform in time gives us

\begin{align}
\tilde{G}_{ik} &= \frac{1}{s} \left[ C_{kpim} q_p q_m + \eta s \delta_{ik} \right]^{-1} \nonumber \\ & = \frac{1}{s} \left[ \frac{1}{\mu q^2 + \eta s}\hat{t}_i \hat{t}_k  + \frac{1}{\left(\lambda  +2\mu \right) q^2+ \eta s}  \hat{q}_i \hat{q}_k \right],
\end{align}

with the last equality holding for an isotropic medium in $2d$. Here $\hat{t}$ is the vector normal to $\hat{q}$.

We invert the spatial Fourier transform, and then the Laplace transform. The result is 

\begin{widetext}
\begin{align}
G_{ik} &= \frac{1}{8 \pi} \left[\frac{1}{\mu} \Gamma{\left(0, \frac{\eta r^2}{4 \mu t}\right)} +  \frac{1}{2 \mu + \lambda}\Gamma{\left(0, \frac{\eta r^2}{4 \left(\lambda + 2\mu \right) t}\right)} + \frac{4 t }{\eta r^2} \left( e^{-\frac{\eta r^2}{4\mu  t } }- e^{-\frac{\eta r^2}{4\left(2 \mu + \lambda\right) t }}  \right) \right] \delta_{ik}  \nonumber \\ &+  \frac{t}{\pi \eta r^2} \left(  e^{-\frac{\eta r^2}{4\left(2 \mu + \lambda\right) t }} - e^{-\frac{\eta r^2}{4\mu  t } } \right) \hat{r}_i \hat{r}_k, 
\end{align}
\end{widetext}
where $\Gamma{\left(0, x\right)} \equiv \int_x^\infty \dd{s} s^{-1} e^{-s} $ is the incomplete gamma function (in this case, also the exponential integral function).

Differentiating this twice and symmetrizing over one of the indices allows us to compute $G_{ijkl}$, the strain response to a dipole of force.

We obtain

\begin{widetext}
\begin{align}
G_{ijkl} &=  \frac{1}{4 \pi r^2} \left[\frac{1}{\mu} e^{-\frac{ \eta r^2}{4 \mu t} }  + \frac{4 t}{\eta r^2} \left( e^{-\frac{\eta r^2}{4\mu  t } }- e^{-\frac{\eta r^2}{4\left(2 \mu + \lambda\right) t }}  \right)\right] \left[\delta_{il} \delta_{jk} + \delta_{jl} \delta_{ik} \right]\\
&- \nonumber \frac{1}{4 \pi r^2} \left[ \frac{16 t}{\eta r^2} \left( e^{-\frac{\eta r^2}{4\mu  t } }- e^{-\frac{\eta r^2}{4\left(2 \mu + \lambda\right) t }}  \right) + \left(\frac{2 }{\mu} e^{-\frac{\eta r^2}{4 \mu t}} -\frac{2 }{2\mu + \lambda} e^{-\frac{\eta r^2}{4 \left(2 \mu+ \lambda\right) t}}  \right)\right]\left[\hat{r}_k \hat{r}_l \delta_{ij}  + \hat{r}_i \hat{r}_j \delta_{kl} + \hat{r}_i \hat{r}_k \delta_{jl} + \hat{r}_j \hat{r}_k \delta_{il}\right] \\
&+ \nonumber \frac{1}{\pi r^2} \left[ \frac{1}{\mu} \left(4 + \frac{\eta r^2}{4 \mu t} + \frac{24 \mu t}{\eta r^2} \right) e^{- \frac{\eta r^2}{4 \mu t}} - \frac{1}{2\mu+ \lambda} \left(4 + \frac{\eta r^2}{4\left(2 \mu + \lambda\right) t} + \frac{24 \left(2\mu + \lambda\right) t}{\eta r^2} \right) e^{- \frac{\eta r^2}{4 \left(2\mu + \lambda\right) t}}  \right] \hat{r}_i \hat{r}_j \hat{r}_k \hat{r}_l \\
&- \nonumber  \frac{1}{4 \pi r^2} \left[  \left(\frac{1}{\mu} + \frac{\eta r^2}{2 \mu t} \right) e^{-\frac{ \eta r^2}{4 \mu t} } - \frac{2}{2 \mu + \lambda} e^{-\frac{ \eta r^2}{4 \left(2 \mu + \lambda \right)  t} } + \frac{16 t}{\eta r^2} \left(e^{-\frac{ \eta r^2}{4 \mu t} } - e^{-\frac{ \eta r^2}{4 \left( \mu + \lambda \right)  t} }  \right) \right]  \nonumber \left[    \hat{r}_j \hat{r}_l \delta_{ik}  + \hat{r}_i \hat{r}_l \delta_{jk}  \right]\\ 
&+   \nonumber   \frac{t}{2 \pi \eta r^4} \left( e^{-\frac{\eta r^2}{4\mu  t } }- e^{-\frac{\eta r^2}{4\left(2 \mu + \lambda\right) t }}  \right) \delta_{ij} \delta_{kl}.
\end{align}
\end{widetext}

Following previous work, a dipole of $xy$ shear strain at the origin is equivalent to a pair of force dipoles \cite{picard2004elastic}.  Assuming this source gives us the elastic strain field (now written in terms of the Poisson ratio $\nu$ and the ``diffusion constants''  $D_\mathrm{T} \equiv \frac{\mu}{\eta} $  and $D_\mathrm{L} \equiv \frac{\lambda + 2 \mu}{\eta} = 2 D_\mathrm{T} / \left(1 - \nu\right) $ 

\begin{align}
k^{\left(\mathrm{S}\right)}{\left(\mathbf{r},t\right)} = - \frac{\left(1 - \nu \right) \sin{2 \theta}}{8 \pi r^2} e^{-\frac{ r^2}{4 D_\mathrm{L}  t}} \left(1 + \frac{ r^2}{4 D_\mathrm{L} t }\right) ,
\label{eq:analyticalK}
\end{align}

\begin{align}
&\epsilon_{xy}^{\left(\mathrm{S}\right)}{\left(\mathbf{r},t\right)} \nonumber \\  &= \frac{ \cos{4 \theta}}{2 \pi r^2}  \ \left[\left(1-\nu\right) e^{-\frac{ r^2}{4 D_\mathrm{L}  t}} \left(2 + \frac{r^2}{8 D_{\mathrm{L}} t} + \frac{12 D_{\mathrm{L}} t}{r^2}\right) \right.  \nonumber \\ &- \left. e^{-\frac{ r^2}{4 D_\mathrm{T}  t}} \left(4 + \frac{r^2}{4 D_\mathrm{T} t} + \frac{24 D_{\mathrm{T}} t}{r^2} \right)  \right]  \nonumber \\ &- \frac{1}{2 \pi r^2}\left( \frac{r^2}{4 D_{\mathrm{L}} t} e^{-\frac{ r^2}{4 D_\mathrm{L}  t}}  + \frac{r^2}{4 D_{\mathrm{T}} t} e^{-\frac{ r^2}{4 D_\mathrm{T}  t}}\right),
\label{eq:analyticalExy}
\end{align}

\begin{align}
&\frac{1}{2}\left(\epsilon_{xx}^{\left(\mathrm{S}\right)}{\left(\mathbf{r}, t\right)} -\epsilon_{yy}^{\left(\mathrm{S}\right)}{\left(\mathbf{r},t\right)}  \right) \nonumber \\ &=  \frac{ \sin{4 \theta}}{2 \pi r^2}  \left[e^{-\frac{ r^2}{4 D_\mathrm{T}  t}} \left(4 + \frac{r^2}{4 D_\mathrm{T} t} + \frac{24 D_{\mathrm{T}} t}{r^2} \right) \right. \nonumber \\ &- \left. \left(1-\nu\right) e^{-\frac{ r^2}{4 D_\mathrm{L}  t}} \left(2 + \frac{r^2}{8 D_{\mathrm{L}} t} + \frac{12 D_{\mathrm{L}} t}{r^2}\right) \right]
\label{eq:analyticalExx}
\end{align}

The familiar power law dependences from elastic equilibrium are realized in the large-time limit

\begin{align}
k^{\left(\mathrm{S}\right)}{\left(\mathbf{r}, \infty\right)} = \frac{\left(\nu - 1 \right) \sin{2 \theta}}{2 \pi r^2} \nonumber \\
\epsilon_{xy}^{\left(\mathrm{S}\right)}{\left(\mathbf{r}, \infty\right)} =  \frac{\left(1 + \nu \right) \cos{4 \theta}}{2 \pi r^2} \nonumber \\
\frac{1}{2} \left(\epsilon_{xx}^{\left(\mathrm{S}\right)}{\left(\mathbf{r}, \infty\right)} -\epsilon_{yy}^{\left(\mathrm{S}\right)}{\left(\mathbf{r},\infty\right)}  \right) =  -\frac{\left(1 + \nu \right) \sin{4 \theta}}{2 \pi r^2} 
\label{eq:epsilon}
\end{align}

These results together explain why the volumetric strain is observed to have a $\sin(2\theta)$ dependence, and why the deviatoric strain magnitude is isotropic. 

Notice, however, that $\int \dd{\theta} k(r,\theta, t) = 0$ for such a shear strain source.  To explain the apparent nonzero value of $\int \dd{\theta} k(r,\theta, t)$ for short times in our simulations, we must consider the effect of a transient expansion source.  The local region surrounding a rearrangement might be expected, on average, to have a different volume than in the initial state. 

In an infinite system, the kernel above gives for a point plastic compression at the origin:

\begin{align}
k^{\left(\mathrm{C}\right)}{\left(\mathbf{r},t\right)} = \frac{1+\nu}{2}  \frac{1}{4 \pi D_{\mathrm{L}} t} e^{- \frac{r^2}{4 D_{\mathrm{L}} t}}.
\label{eq:kr}
\end{align}

As long as the Poisson ratio is close to 1, this precisely conserves volume in an infinite system, when added to the point compression at the origin. 

We expect that since our system is finite (and the short-time Poisson ratio is far from $1$), this kernel would need to be modified near the boundaries of the system to satisfy the periodic boundary conditions and conserve the total volume. We find that it works adequately for the bulk for our data however, and our data at $r$ close to the box size are difficult to resolve - we have chosen the $y$-range in the top-left box of Fig.~\ref{fig:strainField} to exclude points beyond $r=30$ because the error bars are comparable to the absolute value.

The full response to a given event will be a sum of the responses to strain [Eq.~(\ref{eq:epsilon})] and compression sources~(\ref{eq:kr}) with appropriate prefactors, although for measurements where its contribution is nonzero we expect the strain source to be dominant.

\section{Compare analytical and numerical $k(r)$, ${\tilde \epsilon}$, and $\epsilon_{xy}$ results}

Since we have derived a analytical formulae for the strain, Eqs.  (\ref{eq:epsilon}) and (\ref{eq:kr}), we can make comparison with our numerical results. We have numerically measured instantaneous elastic constants $\lambda + 2 \mu=0.3533$ and $\nu=0.3408$ for our system by applying a small ($\ee{-6}$) strain on the simulation box and measuring the force. 

The time interval between frames, $t$, is not fixed since we record frames that are equidistant in configuration space; see the SI \cite{supplement}. We plot the distribution of times between frames in supplementary Fig.~S1, and find that the most probable time interval is $t \approx 100$. The definition of our time implies that $\eta=1$. With these parameters, Eq. (\ref{eq:kr}) predicts a Gaussian that decays to $0.1\%$ of its peak height at $r=31$, roughly consistent with the actual result presented in Fig.~\ref{fig:strainField}.

For the total deviatoric strain ${\tilde \epsilon}$ and $xy$-strain $\epsilon_{xy}$, we have numerically confirmed that they decay as power laws: ${\tilde \epsilon}={\tilde c}/r^2$ and $\epsilon_{xy}=c_{xy}/r^2$ (Fig.~\ref{fig:strainField}), which matches the prediction in Eq.~(\ref{eq:epsilon}). The prefactors, {\it i.e.,} constants ${\tilde c}$ and $c_{xy}$, were not predicted in Appendix~\ref{sec:prediction} since our theory does not take into consideration the average amount of plastic strain caused by a rearranger.

Nevertheless, we can approximately measure this quantity. The strains in equation (\ref{eq:epsilon}) are for a plastic strain $\epsilon_{xy}^{\mathrm{pl}} = \epsilon_0 A \delta{\left(\mathbf{r}\right)}$, i.e. the prefactor of the far-field strain is equal to the product of the area $A = \pi r_\mathrm{D}^2$ of the rearrangement and its plastic strain.

If the rearrangements have a distribution of plastic strains $\epsilon_0$ and orientations $\theta'$, then by rotating the kernel and assuming the distribution of $\theta'$ is even we find that 

\begin{align}
    \langle\tilde{\epsilon}{\left(r\right)}\rangle &= \frac{1+\nu}{2} \frac{r_{\mathrm{D}}^2 \langle \tilde{\epsilon_0} \rangle}{r^2} \\
    \langle\epsilon_{xy}{\left(r\right)}\rangle &= \frac{1+\nu}{2} r_{\mathrm{D}}^2 \langle \epsilon_{0,xy} \cos(4\theta')\rangle \frac{\cos(4\theta)}{r^2} \\
        \langle k{\left(r\right)}\rangle &= \frac{\nu - 1}{2} r_{\mathrm{D}}^2 \langle \epsilon_{0,xy} \cos(2 \theta')\rangle \frac{\sin(2\theta)}{r^2}
\end{align}

We will neglect the $\cos{4\theta'}$ and $\cos{2 \theta'}$ in our rough estimates. 

We find that the $\dtwomin$ correlation length \cite{cubuk2018Science} is $r_D=3.6$, {\it i.e.,} the correlation between $\dtwomin(\mathbf 0)$ and  $\dtwomin(\mathbf r)$ is approximately $\exp(-|\mathbf r|/3.6)$ for small $|\mathbf r|$. The area of the event is then estimated as $\pi r_D^2$. We then calculate the local-fit deviatoric and $xy$-strain within a radius of $r_D$ around each rearranger, and find on average ${\tilde \epsilon}=3.6\e{-3}$ and $\epsilon_{xy}=1.8\e{-4}$ at the rearranging site.  Theoretically, this predicts that the prefactors are ${\tilde c}= \frac{1+\nu}{2}{\tilde \epsilon}r_D^2=0.031$, $c_{xy}=\frac{1+\nu}{2}\epsilon_{xy}r_D^2 \langle\left| \cos{4\theta}\right| \rangle=1.0\e{-3}$, and $c_k = \frac{1 - \nu}{2}\epsilon_{xy}r_D^2 \langle\left| \sin{2\theta}\right|\rangle = 5.0\e{-4}$. This roughly matches the fits presented in Fig.~2 of the main text, which have ${\tilde c}=0.03$, $c_{xy}=1.5\e{-3}$, and $c_k=5.0\e{-4}$.

Why do our numerical results match the analytical derivations for shear strains produced by a shear source, Eqs.~(\ref{eq:analyticalExy})~and~(\ref{eq:analyticalExx}), in the infinite-time limit of Eq.~(\ref{eq:epsilon}), but match that for the volumetric strain produced by a compression source, Eq.~(\ref{eq:kr}), at a finite time? It turns out that at the rearranging site, the plastic shear occurs over a much longer time interval than the plastic compression. We plot these strain components at the rearranging site versus time in supplementary Fig.~S5.
If we approximate such strain-time curves with Gaussians, then the numerically measured strain at distance $r$ should be the convolution of previously-derived finite-time analytical result and Gaussians, {\it i.e.,}
\begin{equation}
\begin{aligned}
    k(r, \mbox{numerical})&=c_k\int_{-\infty}^0 \exp(-\alpha t^2)k(r, t-t')dt',\\
    \epsilon_{xy}(r, \mbox{numerical})&=c_{xy}\int_{-\infty}^0 \exp(-\beta t^2)\epsilon_{xy}(r, t-t')dt',
\end{aligned}
\label{eq:conv}
\end{equation}
where $k(r, t)$ and $\epsilon_{xy}(r, t)$ are given in Eqs.~(\ref{eq:kr})~and~(\ref{eq:analyticalExy}), respectively. We numerically compute these integrals for various parameters. For $k$, the integral fits numerical data well at $\alpha=6.1197\e{-5}$, as shown in Fig.~\ref{fig:strainField}. This indicates that the width of the Gaussian is about $\alpha^{-1/2}=127.83$, roughly consistent with supplementary Fig.~S5. For $\epsilon_{xy}$, however, it turns out that Eq.~(\ref{eq:conv}) cannot closely fit our numerical result, which decays slightly {\it slower} than $r^{-2}$ (Fig.~\ref{fig:strainField}). No matter how small $\beta$ is, Eq.~(\ref{eq:conv}) gives an $\epsilon_{xy}$ that decays slightly {\it faster} than $r^{-2}$. We see two possible reasons for this difference: (1) A finite size effect as $r$ becomes comparable to the box size, or (2) the interference between simultaneous rearrangements in our numerical results. As we discuss in the last paragraph of supplementary Sec. III, we filter out frames with multiple rearrangements, but such filtration cannot be perfect.


\begin{thebibliography}{29}%
\makeatletter
\providecommand \@ifxundefined [1]{%
 \@ifx{#1\undefined}
}%
\providecommand \@ifnum [1]{%
 \ifnum #1\expandafter \@firstoftwo
 \else \expandafter \@secondoftwo
 \fi
}%
\providecommand \@ifx [1]{%
 \ifx #1\expandafter \@firstoftwo
 \else \expandafter \@secondoftwo
 \fi
}%
\providecommand \natexlab [1]{#1}%
\providecommand \enquote  [1]{``#1''}%
\providecommand \bibnamefont  [1]{#1}%
\providecommand \bibfnamefont [1]{#1}%
\providecommand \citenamefont [1]{#1}%
\providecommand \href@noop [0]{\@secondoftwo}%
\providecommand \href [0]{\begingroup \@sanitize@url \@href}%
\providecommand \@href[1]{\@@startlink{#1}\@@href}%
\providecommand \@@href[1]{\endgroup#1\@@endlink}%
\providecommand \@sanitize@url [0]{\catcode `\\12\catcode `\$12\catcode
  `\&12\catcode `\#12\catcode `\^12\catcode `\_12\catcode `\%12\relax}%
\providecommand \@@startlink[1]{}%
\providecommand \@@endlink[0]{}%
\providecommand \url  [0]{\begingroup\@sanitize@url \@url }%
\providecommand \@url [1]{\endgroup\@href {#1}{\urlprefix }}%
\providecommand \urlprefix  [0]{URL }%
\providecommand \Eprint [0]{\href }%
\providecommand \doibase [0]{https://doi.org/}%
\providecommand \selectlanguage [0]{\@gobble}%
\providecommand \bibinfo  [0]{\@secondoftwo}%
\providecommand \bibfield  [0]{\@secondoftwo}%
\providecommand \translation [1]{[#1]}%
\providecommand \BibitemOpen [0]{}%
\providecommand \bibitemStop [0]{}%
\providecommand \bibitemNoStop [0]{.\EOS\space}%
\providecommand \EOS [0]{\spacefactor3000\relax}%
\providecommand \BibitemShut  [1]{\csname bibitem#1\endcsname}%
\let\auto@bib@innerbib\@empty
\bibitem [{\citenamefont {Cubuk}\ \emph {et~al.}(2017)\citenamefont {Cubuk},
  \citenamefont {Ivancic}, \citenamefont {Schoenholz}, \citenamefont
  {Strickland}, \citenamefont {Basu}, \citenamefont {Davidson}, \citenamefont
  {Fontaine}, \citenamefont {Hor}, \citenamefont {Huang}, \citenamefont {Jiang}
  \emph {et~al.}}]{cubuk2018Science}%
  \BibitemOpen
  \bibfield  {author} {\bibinfo {author} {\bibfnamefont {E.~D.}\ \bibnamefont
  {Cubuk}}, \bibinfo {author} {\bibfnamefont {R.}~\bibnamefont {Ivancic}},
  \bibinfo {author} {\bibfnamefont {S.~S.}\ \bibnamefont {Schoenholz}},
  \bibinfo {author} {\bibfnamefont {D.}~\bibnamefont {Strickland}}, \bibinfo
  {author} {\bibfnamefont {A.}~\bibnamefont {Basu}}, \bibinfo {author}
  {\bibfnamefont {Z.}~\bibnamefont {Davidson}}, \bibinfo {author}
  {\bibfnamefont {J.}~\bibnamefont {Fontaine}}, \bibinfo {author}
  {\bibfnamefont {J.~L.}\ \bibnamefont {Hor}}, \bibinfo {author} {\bibfnamefont
  {Y.-R.}\ \bibnamefont {Huang}}, \bibinfo {author} {\bibfnamefont
  {Y.}~\bibnamefont {Jiang}}, \emph {et~al.},\ }\bibfield  {title} {\bibinfo
  {title} {Structure-property relationships from universal signatures of
  plasticity in disordered solids},\ }\href@noop {} {\bibfield  {journal}
  {\bibinfo  {journal} {Science}\ }\textbf {\bibinfo {volume} {358}},\ \bibinfo
  {pages} {1033} (\bibinfo {year} {2017})}\BibitemShut {NoStop}%
\bibitem [{\citenamefont {Durian}(1995)}]{durian1995foam}%
  \BibitemOpen
  \bibfield  {author} {\bibinfo {author} {\bibfnamefont {D.~J.}\ \bibnamefont
  {Durian}},\ }\bibfield  {title} {\bibinfo {title} {Foam mechanics at the
  bubble scale},\ }\href@noop {} {\bibfield  {journal} {\bibinfo  {journal}
  {Phys. Rev. Lett.}\ }\textbf {\bibinfo {volume} {75}},\ \bibinfo {pages}
  {4780} (\bibinfo {year} {1995})}\BibitemShut {NoStop}%
\bibitem [{\citenamefont {Sethna}\ \emph {et~al.}(2001)\citenamefont {Sethna},
  \citenamefont {Dahmen},\ and\ \citenamefont {Myers}}]{sethna2001crackling}%
  \BibitemOpen
  \bibfield  {author} {\bibinfo {author} {\bibfnamefont {J.~P.}\ \bibnamefont
  {Sethna}}, \bibinfo {author} {\bibfnamefont {K.~A.}\ \bibnamefont {Dahmen}},\
  and\ \bibinfo {author} {\bibfnamefont {C.~R.}\ \bibnamefont {Myers}},\
  }\bibfield  {title} {\bibinfo {title} {Crackling noise},\ }\href@noop {}
  {\bibfield  {journal} {\bibinfo  {journal} {Nature}\ }\textbf {\bibinfo
  {volume} {410}},\ \bibinfo {pages} {242} (\bibinfo {year}
  {2001})}\BibitemShut {NoStop}%
\bibitem [{\citenamefont {Salje}\ and\ \citenamefont
  {Dahmen}(2014)}]{saljeDahmen2014crackling}%
  \BibitemOpen
  \bibfield  {author} {\bibinfo {author} {\bibfnamefont {E.~K.~H.}\
  \bibnamefont {Salje}}\ and\ \bibinfo {author} {\bibfnamefont {K.~A.}\
  \bibnamefont {Dahmen}},\ }\bibfield  {title} {\bibinfo {title} {Crackling
  noise in disordered mateirals},\ }\href@noop {} {\bibfield  {journal}
  {\bibinfo  {journal} {Annu. Rev. Condens. Matter Phys.}\ }\textbf {\bibinfo
  {volume} {5}},\ \bibinfo {pages} {233} (\bibinfo {year} {2014})}\BibitemShut
  {NoStop}%
\bibitem [{\citenamefont {Sethna}\ \emph {et~al.}(2017)\citenamefont {Sethna},
  \citenamefont {Bierbaum}, \citenamefont {Dahmen}, \citenamefont {Goodrich},
  \citenamefont {Greer}, \citenamefont {Hayden}, \citenamefont {Kent-Dobias},
  \citenamefont {Lee}, \citenamefont {Liarte}, \citenamefont {Ni} \emph
  {et~al.}}]{sethna2017deformation}%
  \BibitemOpen
  \bibfield  {author} {\bibinfo {author} {\bibfnamefont {J.~P.}\ \bibnamefont
  {Sethna}}, \bibinfo {author} {\bibfnamefont {M.~K.}\ \bibnamefont
  {Bierbaum}}, \bibinfo {author} {\bibfnamefont {K.~A.}\ \bibnamefont
  {Dahmen}}, \bibinfo {author} {\bibfnamefont {C.~P.}\ \bibnamefont
  {Goodrich}}, \bibinfo {author} {\bibfnamefont {J.~R.}\ \bibnamefont {Greer}},
  \bibinfo {author} {\bibfnamefont {L.~X.}\ \bibnamefont {Hayden}}, \bibinfo
  {author} {\bibfnamefont {J.~P.}\ \bibnamefont {Kent-Dobias}}, \bibinfo
  {author} {\bibfnamefont {E.~D.}\ \bibnamefont {Lee}}, \bibinfo {author}
  {\bibfnamefont {D.~B.}\ \bibnamefont {Liarte}}, \bibinfo {author}
  {\bibfnamefont {X.}~\bibnamefont {Ni}}, \emph {et~al.},\ }\bibfield  {title}
  {\bibinfo {title} {Deformation of crystals: Connections with statistical
  physics},\ }\href@noop {} {\bibfield  {journal} {\bibinfo  {journal} {Annu.
  Rev. Mater. Res.}\ }\textbf {\bibinfo {volume} {47}},\ \bibinfo {pages} {217}
  (\bibinfo {year} {2017})}\BibitemShut {NoStop}%
\bibitem [{\citenamefont {Dahmen}\ \emph {et~al.}(2019)\citenamefont {Dahmen},
  \citenamefont {Uhl},\ and\ \citenamefont
  {Wright}}]{dahmen2019FrontiersinPhysicsreview}%
  \BibitemOpen
  \bibfield  {author} {\bibinfo {author} {\bibfnamefont {K.~A.}\ \bibnamefont
  {Dahmen}}, \bibinfo {author} {\bibfnamefont {J.~T.}\ \bibnamefont {Uhl}},\
  and\ \bibinfo {author} {\bibfnamefont {W.~J.}\ \bibnamefont {Wright}},\
  }\bibfield  {title} {\bibinfo {title} {Why the crackling deformations of
  single crystals, metallic glasses, rock, granular materials, and the
  earth’s crust are so surprisingly similar},\ }\href@noop {} {\bibfield
  {journal} {\bibinfo  {journal} {Front. Phys.}\ }\textbf {\bibinfo {volume}
  {7}},\ \bibinfo {pages} {176} (\bibinfo {year} {2019})}\BibitemShut {NoStop}%
\bibitem [{\citenamefont {Conner}\ \emph {et~al.}(2003)\citenamefont {Conner},
  \citenamefont {Johnson}, \citenamefont {Paton},\ and\ \citenamefont
  {Nix}}]{conner2003shear}%
  \BibitemOpen
  \bibfield  {author} {\bibinfo {author} {\bibfnamefont {R.}~\bibnamefont
  {Conner}}, \bibinfo {author} {\bibfnamefont {W.~L.}\ \bibnamefont {Johnson}},
  \bibinfo {author} {\bibfnamefont {N.}~\bibnamefont {Paton}},\ and\ \bibinfo
  {author} {\bibfnamefont {W.}~\bibnamefont {Nix}},\ }\bibfield  {title}
  {\bibinfo {title} {Shear bands and cracking of metallic glass plates in
  bending},\ }\href@noop {} {\bibfield  {journal} {\bibinfo  {journal} {J.
  Appl. Phys.}\ }\textbf {\bibinfo {volume} {94}},\ \bibinfo {pages} {904}
  (\bibinfo {year} {2003})}\BibitemShut {NoStop}%
\bibitem [{\citenamefont {Nicolas}\ \emph {et~al.}(2018)\citenamefont
  {Nicolas}, \citenamefont {Ferrero}, \citenamefont {Martens},\ and\
  \citenamefont {Barrat}}]{nicolas2017deformation}%
  \BibitemOpen
  \bibfield  {author} {\bibinfo {author} {\bibfnamefont {A.}~\bibnamefont
  {Nicolas}}, \bibinfo {author} {\bibfnamefont {E.~E.}\ \bibnamefont
  {Ferrero}}, \bibinfo {author} {\bibfnamefont {K.}~\bibnamefont {Martens}},\
  and\ \bibinfo {author} {\bibfnamefont {J.-L.}\ \bibnamefont {Barrat}},\
  }\bibfield  {title} {\bibinfo {title} {Deformation and flow of amorphous
  solids: Insights from elastoplastic models},\ }\href@noop {} {\bibfield
  {journal} {\bibinfo  {journal} {Rev. Mod. Phys.}\ }\textbf {\bibinfo {volume}
  {90}},\ \bibinfo {pages} {045006} (\bibinfo {year} {2018})}\BibitemShut
  {NoStop}%
\bibitem [{\citenamefont {Shavit}\ and\ \citenamefont
  {Riggleman}(2014)}]{shavitriggleman2014PhysChemChemPhys}%
  \BibitemOpen
  \bibfield  {author} {\bibinfo {author} {\bibfnamefont {A.}~\bibnamefont
  {Shavit}}\ and\ \bibinfo {author} {\bibfnamefont {R.~A.}\ \bibnamefont
  {Riggleman}},\ }\bibfield  {title} {\bibinfo {title} {Strain localization in
  glassy polymers under cylindrical confinement},\ }\href@noop {} {\bibfield
  {journal} {\bibinfo  {journal} {Phys. Chem. Chem. Phys.}\ }\textbf {\bibinfo
  {volume} {16}},\ \bibinfo {pages} {10301} (\bibinfo {year}
  {2014})}\BibitemShut {NoStop}%
\bibitem [{\citenamefont {Ozawa}\ \emph {et~al.}(2018)\citenamefont {Ozawa},
  \citenamefont {Berthier}, \citenamefont {Biroli}, \citenamefont {Rosso},\
  and\ \citenamefont {Tarjus}}]{ozawa2018pnas}%
  \BibitemOpen
  \bibfield  {author} {\bibinfo {author} {\bibfnamefont {M.}~\bibnamefont
  {Ozawa}}, \bibinfo {author} {\bibfnamefont {L.}~\bibnamefont {Berthier}},
  \bibinfo {author} {\bibfnamefont {G.}~\bibnamefont {Biroli}}, \bibinfo
  {author} {\bibfnamefont {A.}~\bibnamefont {Rosso}},\ and\ \bibinfo {author}
  {\bibfnamefont {G.}~\bibnamefont {Tarjus}},\ }\bibfield  {title} {\bibinfo
  {title} {Random critical point separates brittle and ductile yielding
  transitions in amorphous materials},\ }\href@noop {} {\bibfield  {journal}
  {\bibinfo  {journal} {Proc. Natl. Acad. Sci.}\ }\textbf {\bibinfo {volume}
  {115}},\ \bibinfo {pages} {6656} (\bibinfo {year} {2018})}\BibitemShut
  {NoStop}%
\bibitem [{\citenamefont {Spaepen}(1977)}]{spaepen1977microscopic}%
  \BibitemOpen
  \bibfield  {author} {\bibinfo {author} {\bibfnamefont {F.}~\bibnamefont
  {Spaepen}},\ }\bibfield  {title} {\bibinfo {title} {A microscopic mechanism
  for steady state inhomogeneous flow in metallic glasses},\ }\href@noop {}
  {\bibfield  {journal} {\bibinfo  {journal} {Acta Metallurgica}\ }\textbf
  {\bibinfo {volume} {25}},\ \bibinfo {pages} {407} (\bibinfo {year}
  {1977})}\BibitemShut {NoStop}%
\bibitem [{\citenamefont {Falk}\ and\ \citenamefont
  {Langer}(2011)}]{falk2011review}%
  \BibitemOpen
  \bibfield  {author} {\bibinfo {author} {\bibfnamefont {M.~L.}\ \bibnamefont
  {Falk}}\ and\ \bibinfo {author} {\bibfnamefont {J.~S.}\ \bibnamefont
  {Langer}},\ }\bibfield  {title} {\bibinfo {title} {Deformation and failure of
  amorphous, solidlike materials},\ }\href@noop {} {\bibfield  {journal}
  {\bibinfo  {journal} {Annu. Rev. Condens. Matter Phys.}\ }\textbf {\bibinfo
  {volume} {2}},\ \bibinfo {pages} {353} (\bibinfo {year} {2011})}\BibitemShut
  {NoStop}%
\bibitem [{\citenamefont {Widmer-Cooper}\ \emph {et~al.}(2008)\citenamefont
  {Widmer-Cooper}, \citenamefont {Perry}, \citenamefont {Harrowell},\ and\
  \citenamefont {Reichman}}]{harrowell2008irreversible}%
  \BibitemOpen
  \bibfield  {author} {\bibinfo {author} {\bibfnamefont {A.}~\bibnamefont
  {Widmer-Cooper}}, \bibinfo {author} {\bibfnamefont {H.}~\bibnamefont
  {Perry}}, \bibinfo {author} {\bibfnamefont {P.}~\bibnamefont {Harrowell}},\
  and\ \bibinfo {author} {\bibfnamefont {D.~R.}\ \bibnamefont {Reichman}},\
  }\bibfield  {title} {\bibinfo {title} {Irreversible reorganization in a
  supercooled liquid originates from localized soft modes},\ }\href@noop {}
  {\bibfield  {journal} {\bibinfo  {journal} {Nat. Phys.}\ }\textbf {\bibinfo
  {volume} {4}},\ \bibinfo {pages} {711} (\bibinfo {year} {2008})}\BibitemShut
  {NoStop}%
\bibitem [{\citenamefont {Manning}\ and\ \citenamefont
  {Liu}(2011)}]{manning2011vibrational}%
  \BibitemOpen
  \bibfield  {author} {\bibinfo {author} {\bibfnamefont {M.~L.}\ \bibnamefont
  {Manning}}\ and\ \bibinfo {author} {\bibfnamefont {A.~J.}\ \bibnamefont
  {Liu}},\ }\bibfield  {title} {\bibinfo {title} {Vibrational modes identify
  soft spots in a sheared disordered packing},\ }\href@noop {} {\bibfield
  {journal} {\bibinfo  {journal} {Phys. Rev. Lett.}\ }\textbf {\bibinfo
  {volume} {107}},\ \bibinfo {pages} {108302} (\bibinfo {year}
  {2011})}\BibitemShut {NoStop}%
\bibitem [{\citenamefont {Schoenholz}\ \emph {et~al.}(2016)\citenamefont
  {Schoenholz}, \citenamefont {Cubuk}, \citenamefont {Sussman}, \citenamefont
  {Kaxiras},\ and\ \citenamefont {Liu}}]{schoenholz2016structural}%
  \BibitemOpen
  \bibfield  {author} {\bibinfo {author} {\bibfnamefont {S.~S.}\ \bibnamefont
  {Schoenholz}}, \bibinfo {author} {\bibfnamefont {E.~D.}\ \bibnamefont
  {Cubuk}}, \bibinfo {author} {\bibfnamefont {D.~M.}\ \bibnamefont {Sussman}},
  \bibinfo {author} {\bibfnamefont {E.}~\bibnamefont {Kaxiras}},\ and\ \bibinfo
  {author} {\bibfnamefont {A.~J.}\ \bibnamefont {Liu}},\ }\bibfield  {title}
  {\bibinfo {title} {A structural approach to relaxation in glassy liquids},\
  }\href@noop {} {\bibfield  {journal} {\bibinfo  {journal} {Nat. Phys.}\
  }\textbf {\bibinfo {volume} {12}},\ \bibinfo {pages} {469} (\bibinfo {year}
  {2016})}\BibitemShut {NoStop}%
\bibitem [{\citenamefont {Schoenholz}\ \emph {et~al.}(2017)\citenamefont
  {Schoenholz}, \citenamefont {Cubuk}, \citenamefont {Kaxiras},\ and\
  \citenamefont {Liu}}]{schoenholz2017relationship}%
  \BibitemOpen
  \bibfield  {author} {\bibinfo {author} {\bibfnamefont {S.~S.}\ \bibnamefont
  {Schoenholz}}, \bibinfo {author} {\bibfnamefont {E.~D.}\ \bibnamefont
  {Cubuk}}, \bibinfo {author} {\bibfnamefont {E.}~\bibnamefont {Kaxiras}},\
  and\ \bibinfo {author} {\bibfnamefont {A.~J.}\ \bibnamefont {Liu}},\
  }\bibfield  {title} {\bibinfo {title} {Relationship between local structure
  and relaxation in out-of-equilibrium glassy systems},\ }\href@noop {}
  {\bibfield  {journal} {\bibinfo  {journal} {Proc. Natl. Acad. Sci.}\ }\textbf
  {\bibinfo {volume} {114}},\ \bibinfo {pages} {263} (\bibinfo {year}
  {2017})}\BibitemShut {NoStop}%
\bibitem [{\citenamefont {Sharp}\ \emph {et~al.}(2018)\citenamefont {Sharp},
  \citenamefont {Thomas}, \citenamefont {Cubuk}, \citenamefont {Schoenholz},
  \citenamefont {Srolovitz},\ and\ \citenamefont {Liu}}]{sharp2018machine}%
  \BibitemOpen
  \bibfield  {author} {\bibinfo {author} {\bibfnamefont {T.~A.}\ \bibnamefont
  {Sharp}}, \bibinfo {author} {\bibfnamefont {S.~L.}\ \bibnamefont {Thomas}},
  \bibinfo {author} {\bibfnamefont {E.~D.}\ \bibnamefont {Cubuk}}, \bibinfo
  {author} {\bibfnamefont {S.~S.}\ \bibnamefont {Schoenholz}}, \bibinfo
  {author} {\bibfnamefont {D.~J.}\ \bibnamefont {Srolovitz}},\ and\ \bibinfo
  {author} {\bibfnamefont {A.~J.}\ \bibnamefont {Liu}},\ }\bibfield  {title}
  {\bibinfo {title} {Machine learning determination of atomic dynamics at grain
  boundaries},\ }\href@noop {} {\bibfield  {journal} {\bibinfo  {journal}
  {Proc. Natl. Acad. Sci.}\ }\textbf {\bibinfo {volume} {115}},\ \bibinfo
  {pages} {10943} (\bibinfo {year} {2018})}\BibitemShut {NoStop}%
\bibitem [{\citenamefont {Sussman}\ \emph {et~al.}(2017)\citenamefont
  {Sussman}, \citenamefont {Schoenholz}, \citenamefont {Cubuk},\ and\
  \citenamefont {Liu}}]{sussman2018films}%
  \BibitemOpen
  \bibfield  {author} {\bibinfo {author} {\bibfnamefont {D.~M.}\ \bibnamefont
  {Sussman}}, \bibinfo {author} {\bibfnamefont {S.~S.}\ \bibnamefont
  {Schoenholz}}, \bibinfo {author} {\bibfnamefont {E.~D.}\ \bibnamefont
  {Cubuk}},\ and\ \bibinfo {author} {\bibfnamefont {A.~J.}\ \bibnamefont
  {Liu}},\ }\bibfield  {title} {\bibinfo {title} {Disconnecting structure and
  dynamics in glassy thin films},\ }\href@noop {} {\bibfield  {journal}
  {\bibinfo  {journal} {Proc. Nat. Acad. Sci.}\ }\textbf {\bibinfo {volume}
  {114}},\ \bibinfo {pages} {10601} (\bibinfo {year} {2017})}\BibitemShut
  {NoStop}%
\bibitem [{sup()}]{supplement}%
  \BibitemOpen
  \href@noop {} {}\bibinfo {note} {See Supplemental Material at [URL will be
  inserted by publisher]}\BibitemShut {NoStop}%
\bibitem [{\citenamefont {Maloney}\ and\ \citenamefont
  {Lema{\^\i}tre}(2006)}]{maloney2006amorphous}%
  \BibitemOpen
  \bibfield  {author} {\bibinfo {author} {\bibfnamefont {C.~E.}\ \bibnamefont
  {Maloney}}\ and\ \bibinfo {author} {\bibfnamefont {A.}~\bibnamefont
  {Lema{\^\i}tre}},\ }\bibfield  {title} {\bibinfo {title} {Amorphous systems
  in athermal, quasistatic shear},\ }\href@noop {} {\bibfield  {journal}
  {\bibinfo  {journal} {Phys. Rev. E}\ }\textbf {\bibinfo {volume} {74}},\
  \bibinfo {pages} {016118} (\bibinfo {year} {2006})}\BibitemShut {NoStop}%
\bibitem [{\citenamefont {Falk}\ and\ \citenamefont
  {Langer}(1998)}]{falk1998dynamics}%
  \BibitemOpen
  \bibfield  {author} {\bibinfo {author} {\bibfnamefont {M.~L.}\ \bibnamefont
  {Falk}}\ and\ \bibinfo {author} {\bibfnamefont {J.~S.}\ \bibnamefont
  {Langer}},\ }\bibfield  {title} {\bibinfo {title} {Dynamics of viscoplastic
  deformation in amorphous solids},\ }\href@noop {} {\bibfield  {journal}
  {\bibinfo  {journal} {Phys. Rev. E}\ }\textbf {\bibinfo {volume} {57}},\
  \bibinfo {pages} {7192} (\bibinfo {year} {1998})}\BibitemShut {NoStop}%
\bibitem [{\citenamefont {Picard}\ \emph {et~al.}(2004)\citenamefont {Picard},
  \citenamefont {Ajdari}, \citenamefont {Lequeux},\ and\ \citenamefont
  {Bocquet}}]{picard2004elastic}%
  \BibitemOpen
  \bibfield  {author} {\bibinfo {author} {\bibfnamefont {G.}~\bibnamefont
  {Picard}}, \bibinfo {author} {\bibfnamefont {A.}~\bibnamefont {Ajdari}},
  \bibinfo {author} {\bibfnamefont {F.}~\bibnamefont {Lequeux}},\ and\ \bibinfo
  {author} {\bibfnamefont {L.}~\bibnamefont {Bocquet}},\ }\bibfield  {title}
  {\bibinfo {title} {Elastic consequences of a single plastic event: A step
  towards the microscopic modeling of the flow of yield stress fluids},\
  }\href@noop {} {\bibfield  {journal} {\bibinfo  {journal} {Eur. Phys. J. E}\
  }\textbf {\bibinfo {volume} {15}},\ \bibinfo {pages} {371} (\bibinfo {year}
  {2004})}\BibitemShut {NoStop}%
\bibitem [{\citenamefont {Jensen}\ \emph {et~al.}(2014)\citenamefont {Jensen},
  \citenamefont {Weitz},\ and\ \citenamefont {Spaepen}}]{jensen2014local}%
  \BibitemOpen
  \bibfield  {author} {\bibinfo {author} {\bibfnamefont {K.}~\bibnamefont
  {Jensen}}, \bibinfo {author} {\bibfnamefont {D.~A.}\ \bibnamefont {Weitz}},\
  and\ \bibinfo {author} {\bibfnamefont {F.}~\bibnamefont {Spaepen}},\
  }\bibfield  {title} {\bibinfo {title} {Local shear transformations in
  deformed and quiescent hard-sphere colloidal glasses},\ }\href@noop {}
  {\bibfield  {journal} {\bibinfo  {journal} {Phys. Rev. E}\ }\textbf {\bibinfo
  {volume} {90}},\ \bibinfo {pages} {042305} (\bibinfo {year}
  {2014})}\BibitemShut {NoStop}%
\bibitem [{\citenamefont {Karmakar}\ \emph {et~al.}(2010)\citenamefont
  {Karmakar}, \citenamefont {Lerner},\ and\ \citenamefont
  {Procaccia}}]{karmakar2010}%
  \BibitemOpen
  \bibfield  {author} {\bibinfo {author} {\bibfnamefont {S.}~\bibnamefont
  {Karmakar}}, \bibinfo {author} {\bibfnamefont {E.}~\bibnamefont {Lerner}},\
  and\ \bibinfo {author} {\bibfnamefont {I.}~\bibnamefont {Procaccia}},\
  }\bibfield  {title} {\bibinfo {title} {Statistical physics of the yielding
  transition in amorphous solids},\ }\href@noop {} {\bibfield  {journal}
  {\bibinfo  {journal} {Phys. Rev. E.}\ }\textbf {\bibinfo {volume} {82}},\
  \bibinfo {pages} {055103(R)} (\bibinfo {year} {2010})}\BibitemShut {NoStop}%
\bibitem [{\citenamefont {Barbot}\ \emph {et~al.}(2018)\citenamefont {Barbot},
  \citenamefont {Lerbinger}, \citenamefont {Hernandez-Garcia}, \citenamefont
  {Garc{\'\i}a-Garc{\'\i}a}, \citenamefont {Falk}, \citenamefont
  {Vandembroucq},\ and\ \citenamefont {Patinet}}]{barbot2018local}%
  \BibitemOpen
  \bibfield  {author} {\bibinfo {author} {\bibfnamefont {A.}~\bibnamefont
  {Barbot}}, \bibinfo {author} {\bibfnamefont {M.}~\bibnamefont {Lerbinger}},
  \bibinfo {author} {\bibfnamefont {A.}~\bibnamefont {Hernandez-Garcia}},
  \bibinfo {author} {\bibfnamefont {R.}~\bibnamefont
  {Garc{\'\i}a-Garc{\'\i}a}}, \bibinfo {author} {\bibfnamefont {M.~L.}\
  \bibnamefont {Falk}}, \bibinfo {author} {\bibfnamefont {D.}~\bibnamefont
  {Vandembroucq}},\ and\ \bibinfo {author} {\bibfnamefont {S.}~\bibnamefont
  {Patinet}},\ }\bibfield  {title} {\bibinfo {title} {Local yield stress
  statistics in model amorphous solids},\ }\href@noop {} {\bibfield  {journal}
  {\bibinfo  {journal} {Phys. Rev. E}\ }\textbf {\bibinfo {volume} {97}},\
  \bibinfo {pages} {033001} (\bibinfo {year} {2018})}\BibitemShut {NoStop}%
\bibitem [{\citenamefont {Ma}\ \emph {et~al.}(2019)\citenamefont {Ma},
  \citenamefont {Davidson}, \citenamefont {Still}, \citenamefont {Ivancic},
  \citenamefont {Schoenholz}, \citenamefont {Liu},\ and\ \citenamefont
  {Yodh}}]{ma2018prl}%
  \BibitemOpen
  \bibfield  {author} {\bibinfo {author} {\bibfnamefont {X.}~\bibnamefont
  {Ma}}, \bibinfo {author} {\bibfnamefont {Z.~S.}\ \bibnamefont {Davidson}},
  \bibinfo {author} {\bibfnamefont {T.}~\bibnamefont {Still}}, \bibinfo
  {author} {\bibfnamefont {R.~J.}\ \bibnamefont {Ivancic}}, \bibinfo {author}
  {\bibfnamefont {S.}~\bibnamefont {Schoenholz}}, \bibinfo {author}
  {\bibfnamefont {A.}~\bibnamefont {Liu}},\ and\ \bibinfo {author}
  {\bibfnamefont {A.}~\bibnamefont {Yodh}},\ }\bibfield  {title} {\bibinfo
  {title} {Heterogeneous activation, local structure, and softness in
  supercooled colloidal liquids},\ }\href@noop {} {\bibfield  {journal}
  {\bibinfo  {journal} {Phys. Rev. Lett.}\ }\textbf {\bibinfo {volume} {122}},\
  \bibinfo {pages} {028001} (\bibinfo {year} {2019})}\BibitemShut {NoStop}%
\bibitem [{\citenamefont {Budrikis}\ and\ \citenamefont
  {Zapperi}(2013)}]{budrikis2013avalanche}%
  \BibitemOpen
  \bibfield  {author} {\bibinfo {author} {\bibfnamefont {Z.}~\bibnamefont
  {Budrikis}}\ and\ \bibinfo {author} {\bibfnamefont {S.}~\bibnamefont
  {Zapperi}},\ }\bibfield  {title} {\bibinfo {title} {Avalanche localization
  and crossover scaling in amorphous plasticity},\ }\href@noop {} {\bibfield
  {journal} {\bibinfo  {journal} {Phys. Rev. E}\ }\textbf {\bibinfo {volume}
  {88}},\ \bibinfo {pages} {062403} (\bibinfo {year} {2013})}\BibitemShut
  {NoStop}%
\bibitem [{\citenamefont {Cubuk}\ \emph {et~al.}(2015)\citenamefont {Cubuk},
  \citenamefont {Schoenholz}, \citenamefont {Rieser}, \citenamefont {Malone},
  \citenamefont {Rottler}, \citenamefont {Durian}, \citenamefont {Kaxiras},\
  and\ \citenamefont {Liu}}]{cubuk2015PRL}%
  \BibitemOpen
  \bibfield  {author} {\bibinfo {author} {\bibfnamefont {E.~D.}\ \bibnamefont
  {Cubuk}}, \bibinfo {author} {\bibfnamefont {S.~S.}\ \bibnamefont
  {Schoenholz}}, \bibinfo {author} {\bibfnamefont {J.~M.}\ \bibnamefont
  {Rieser}}, \bibinfo {author} {\bibfnamefont {B.~D.}\ \bibnamefont {Malone}},
  \bibinfo {author} {\bibfnamefont {J.}~\bibnamefont {Rottler}}, \bibinfo
  {author} {\bibfnamefont {D.~J.}\ \bibnamefont {Durian}}, \bibinfo {author}
  {\bibfnamefont {E.}~\bibnamefont {Kaxiras}},\ and\ \bibinfo {author}
  {\bibfnamefont {A.~J.}\ \bibnamefont {Liu}},\ }\bibfield  {title} {\bibinfo
  {title} {Identifying structural flow defects in disordered solids using
  machine-learning methods},\ }\href@noop {} {\bibfield  {journal} {\bibinfo
  {journal} {Phys. Rev. Lett.}\ }\textbf {\bibinfo {volume} {114}},\ \bibinfo
  {pages} {108001} (\bibinfo {year} {2015})}\BibitemShut {NoStop}%
\bibitem [{\citenamefont {Harrington}\ \emph {et~al.}(2019)\citenamefont
  {Harrington}, \citenamefont {Liu},\ and\ \citenamefont
  {Durian}}]{harrington2019pre}%
  \BibitemOpen
  \bibfield  {author} {\bibinfo {author} {\bibfnamefont {M.}~\bibnamefont
  {Harrington}}, \bibinfo {author} {\bibfnamefont {A.~J.}\ \bibnamefont
  {Liu}},\ and\ \bibinfo {author} {\bibfnamefont {D.~J.}\ \bibnamefont
  {Durian}},\ }\bibfield  {title} {\bibinfo {title} {Machine learning
  characterization of structural defects in amorphous packings of dimers and
  ellipses},\ }\href@noop {} {\bibfield  {journal} {\bibinfo  {journal} {Phys.
  Rev. E}\ }\textbf {\bibinfo {volume} {99}},\ \bibinfo {pages} {022903}
  (\bibinfo {year} {2019})}\BibitemShut {NoStop}%
\end{thebibliography}
\end{document}